\documentclass[11pt]{article}
\usepackage{amsmath,amssymb, mcite}
\usepackage{rotating}
\usepackage{caption}

\renewcommand{\d}{\textup{d}}

\def\eo{\overset{_{\phantom{.}\circ}}{e}{}}

\def\go{\overset{_{\phantom{.}\circ}}{g}{}}

\def\Do{\overset{_{\phantom{.}\circ}}{D}{}}

\def\So{\overset{_{\phantom{.}\circ}}{S}{}}

\def\etao{\overset{_{\phantom{.}\circ}}{\eta}{}}
\def\zetao{\overset{_{\phantom{.}\circ}}{\zeta}{}}

\newcommand{\tM}{{\tt M}}
\newcommand{\tN}{{\tt N}}

\newcommand{\cV}{{\cal V}}
\newcommand{\cA}{{\cal A}}

\newcommand{\cD}{{\cal D}}
\newcommand{\cQ}{{\cal Q}}
\newcommand{\cP}{{\cal P}}
\newcommand{\cI}{{\cal I}}
\newcommand{\cJ}{{\cal J}}
\newcommand{\cK}{{\cal K}}

\newcommand{\cY}{{\cal Y}}

\begin{document}

\begin{titlepage}

\hfill AEI-2013-240
\vspace{2.5cm}
\begin{center}

{{\LARGE  \bf  Non-linear Kaluza-Klein theory for dual fields}} \\

\vskip 1.5cm {Hadi Godazgar, Mahdi Godazgar
and Hermann Nicolai}
\\
{\vskip 0.8cm
Max-Planck-Institut f\"{u}r Gravitationsphysik, \\
Albert-Einstein-Institut,\\
Am M\"{u}hlenberg 1, D-14476 Potsdam, Germany}
{\vskip 0.4cm
\texttt{Hadi.Godazgar@aei.mpg.de, Mahdi.Godazgar@aei.mpg.de,\\
Hermann.Nicolai@aei.mpg.de}}
\end{center}

\vskip 0.35cm

\begin{center}
\today
\end{center}

\noindent

\vskip 1.2cm

\begin{abstract}
\noindent We present non-linear uplift ans\"atze for all the bosonic degrees of freedom and dual fields in the $S^7$ reduction of $D=11$ supergravity to maximal SO$(8)$ gauged supergravity and test them for the SO$(7)^{\pm}$ invariant solutions. In particular, we complete the known ans\"atze for the internal components of the metric and 4-form flux by constructing a non-linear ansatz for the internal components of the dual 7-form flux. Furthermore, we provide ans\"atze for the complete set of 56 vector fields, which are given by more general structures than those 
available from standard Kaluza-Klein theory. The novel features encountered here have no conventional geometric interpretation and provide a new perspective on Kaluza-Klein theory.
We study the recently found set of generalised vielbein postulates and, for the $S^7$ compactification, we show that they reduce to the E$_{7(7)}$ Cartan equation of 
maximal SO(8) gauged supergravity in four dimensions. The significance of this framework 
for a higher-dimensional understanding of the embedding tensor and other gauged
maximal supergravities is briefly discussed. 

\end{abstract}

\end{titlepage}

\section{Introduction}

In this article we continue the investigation of the generalised geometry 
underlying maximal supergravity on the basis of the SO(1,3) $\!\times\!$ SU(8) invariant 
reformulation \cite{dWNsu8} of $D=11$ supergravity, which has been further developed  in
very recent work \cite{dWN13,GGN,GGN13}. Our main concern here will be the 
question of whether the consistency results of standard Kaluza-Klein theory can be 
extended to the `non-geometrical' structures that arise in this context, and in 
particular from the vector and form fields {\em and their duals}.  Focusing
on $D=11$ supergravity \cite{CJS}, and more specifically on its $S^7$ compactification 
\cite{freundrubin, duffpope} to maximal gauged SO(8) supergravity in four dimensions \cite{dWNn8}
is amply justified by the fact that this theory presents by far the richest 
structure of all Kaluza-Klein models studied so far, but of course, we expect 
our results to be relevant also in a more general context.

The question of whether a four-dimensional theory can be obtained by
dimensional reduction from higher dimensions, and the question of whether 
a given compactification of a higher-dimensional theory can be associated 
with a {\em consistent truncation} is clearly an important and pertinent one.  
Consistency here by definition is taken to mean that any full solution 
of the lower-dimensional theory should admit an uplift to a full solution of the 
non-linear higher-dimensional field equations.  However, establishing such a 
relation and its consistency is far from obvious in all but the most trivial examples.  
In particular, due to the generic emergence of non-abelian gauge theories in 
Kaluza-Klein compactifications, we have to deal with {\em gauged} supergravities.
The most efficient framework for understanding these theories (in any dimension) is the embedding tensor formalism \cite{NSmaximal3, NScomgauge3, dWSTlag, dWSTmax4}.  Therefore, any general scheme that aims to address the issue of the higher-dimensional origin of four-dimensional theories should provide a higher-dimensional perspective on the embedding tensor.  Furthermore, given a consistent truncation, yet another challenging task is to 
give explicit uplift ans\"atze for all relevant fields, something that standard Kaluza-Klein 
theory cannot give for fields other than the graviton and the vector fields.

In \cite{dWNsu8}, a $4+7$ splitting of $D=11$ supergravity is considered with an appropriate decomposition of all eleven-dimensional fields with respect to this splitting, while retaining full on-shell equivalence to the original theory.  This reformulation has manifest local SU$(8)$ 
invariance, and emphasises and generalises the structures that would appear upon a 
toroidal reduction of the theory to four dimensions \cite{cremmerjulia, so(8)}. The construction of \cite{dWNsu8} relies on an analysis of the supersymmetry transformations of the redefined fields and a crucial object that emerges from the supersymmetry transformation of the graviphoton is an SU$(8)$ tensor, the generalised vielbein.  The graviphoton gives rise to vector fields upon reduction. However, in the reduced theory these are complemented by other vector fields.  In particular, the three-form potential also contributes to the vector degrees of freedom. The supersymmetry transformation of these vectors in the $D=11$ theory gives rise to yet another 
generalised vielbein \cite{dWN13}. The observation made in \cite{GGN13} is that by considering dualisation of eleven-dimensional fields, a full set of 56 vectors is obtained whose supersymmetry transformations give rise naturally to an E$_{7(7)}$ vielbein in eleven dimensions. 

The emergence \cite{GGN13} of E$_{7(7)}$ structures in $D=11$ supergravity 
gives a new perspective on the extent to which duality symmetries play a role 
in the full unreduced $D=11$ theory and the necessity to transcend usual notions 
of geometry.~\footnote{For alternative approaches to generalised 
 geometry and a list of recent references with bibliographies, 
 see \cite{Coimbra:2012af, *Park:2013gaj, *Aldazabal:2013mya, *Cederwall:2013naa, *GGPE8, *Hatsuda:2013dya, *Berman:2013eva,  *Hohm:2013pua}.}   
However, the framework that is developed in \cite{GGN13}---based on the SU$(8)$ invariant reformulation \cite{dWNsu8} of $D=11$ supergravity and extending the recent results 
of \cite{dWN13}---is also the most natural setting in which to understand the eleven-dimensional origins of four-dimensional gauged theories.  
The construction in \cite{GGN13} highlights structures in eleven dimensions that are manifest in the reduced theory, enabling one to address questions concerning uplift ans\"atze and the appearance of particular gaugings in four dimensions from a reduction point of view. 

In this paper, we use the $S^7$ reduction to maximal gauged supergravity to illustrate the effectiveness of the framework presented in \cite{GGN13}.  In particular, we extend the Kaluza-Klein ans\"atze for the graviphoton \cite{kerner,chofreund,salamstrathdeeKK} and the vector associated with the three-form potential \cite{dWN13} to vectors associated with dual fields, revealing novel features.  These ans\"atze allow us to derive a non-linear ansatz for the internal components of the six-form potential complementing the non-linear ans\"atze for the internal components of the metric \cite{dWNW} and the three-form potential \cite{dWN13}. A remarkable feature of this analysis is not only the existence of a non-linear ansatz for a \emph{dual field}, but for \emph{all} fields.

To illustrate the novelty of the present construction, 
let us first recall some well-known facts from standard Kaluza-Klein 
theory \cite{kerner,chofreund,salamstrathdeeKK}. Starting from the 
higher-dimensional vielbein $E_M{}^A(z) \equiv E_M{}^A(x,y)$,  where 
the higher-dimensional coordinates $\{z^M\}$ are split into four-dimensional
coordinates $\{x^\mu\}$ and internal coordinates $\{y^m \}$, respectively,
one proceeds from the ansatz
\begin{equation}\label{EMA}
E_M{}^A (x,y) \, = \,
\left(\begin{matrix} \Delta^{-1/2} e'_\mu{}^\alpha  & B_\mu{}^m e_m{}^a \\[2mm]
0 & e_m{}^a \end{matrix}\right),
\end{equation}
where $e_m{}^a(x,y)$ is the vielbein associated to the internal manifold on which the
higher-dimensional theory is assumed to be compactified,  $\Delta\equiv \det e_m{}^a$ 
and $e'_\mu{}^\alpha$ is the Weyl rescaled vierbein of the compactified theory;
the triangular form of $E_M{}^A$ is arrived at by making partial use of the
local Lorentz symmetry of the higher-dimensional theory. A consistent truncation 
for the spin-two field (graviton)  is achieved simply by setting 
$$
e'_\mu{}^\alpha(x,y) \equiv e'_\mu{}^\alpha(x),
$$
that is, by dropping all dependence on the internal coordinates. Likewise, for 
the vectors $B_\mu{}^m(x,y)$, the exact consistent ansatz has been 
known for a very long time \cite{kerner, chofreund, salamstrathdeeKK}; it reads
\begin{equation} 
B_\mu{}^m  (x,y) = K^{m\cI}(y) A_\mu^{\cI}(x),
\end{equation}
where the index $\cI$ labels the Killing vectors $K^{m\cI}(y)$ on the internal manifold.
It is a key result of Kaluza-Klein theory that the non-abelian gauge interactions of the 
compactified  theory then originate from the commutator of two Killing vector fields
\begin{equation}
\big[ K^{m\cI}\partial_m \, , \, K^{n\cJ} \partial_n \big] \, = \, 
f^{\cI\cJ}{}_\cK \, K^{p\cK} \partial_p,
\end{equation}
where $f^{\cI\cJ}{}_\cK$ are the structure constants of the isometry group
of the internal manifold. In this way the gauge group of the compactified theory
is completely explained in geometric terms. While this has been well 
understood for many decades, the main difficulty in establishing the full
consistency of the Kaluza-Klein reduction resides in the scalar sector, in particular
involving the search for consistent ans\"atze for the internal vielbein $e_m{}^a(x,y)$
and other fields of a tensorial nature under internal symmetries. 
The main focus of the present work, then, is to develop a similar theory for `non-geometrical' 
vector fields and matter fields, and in particular for those fields arising from dual fields in 
higher dimensions, for which no readily applicable formulae are available from 
general Kaluza-Klein theory---hence the need for a ``generalised geometry"!

The structure of the paper is as follows.  In section \ref{sec:E711d}, we review the main results of Ref.~\cite{GGN13}.  We briefly discuss how the consideration of dual fields in eleven dimensions can be used to construct 56 vectors, the supersymmetry transformations of which give rise to a set of generalised vielbeine that are parametrised by the `internal' components of the eleven-dimensional metric, three-form potential and its dual six-form.  The generalised vielbeine can be viewed as components of an E$_{7(7)}$ matrix and the supersymmetry transformations of the bosonic fields can be cast into a form that mirrors the analogous supersymmetry transformations in four dimensions.  Furthermore, the generalised vielbein postulates \cite{dWNsu8, GGN13} are summarised in section \ref{subsec:gvp}.

Specialising to the $S^7$ reduction of $D=11$ supergravity to maximal SO$(8)$ gauged supergravity in four dimensions in section \ref{sec:NLA}, we use the results summarised in section \ref{sec:E711d} to derive non-linear ans\"atze for all bosonic degrees of freedom.  
In particular, extending the result in Ref.~\cite{dWN13}, we give Kaluza-Klein ans\"atze 
for all 56 vector fields.  These ans\"atze not only include Killing vector fields on $S^7$,
but also tensors and the potential for the volume form on $S^7$, which is not globally 
defined. Comparing the eleven and four-dimensional supersymmetry transformations 
of the vectors, along the lines of \cite{dWNW, dWN13}, allows us to express the generalised vielbeine in terms of the four-dimensional scalars.  These relations are then used to find 
a non-linear ansatz for the internal components of the six-form dual potential.  The
set of ans\"atze for the vectors and the internal components of the metric, three-form 
potential and its dual comprise a set of uplift formulae for all bosonic degrees of freedom. In section \ref{sec:test}, we test the non-linear ansatz for the 6-form field by explicitly checking 
that the ansatz reproduces the internal component of the 6-form potential of the 
SO$(7)^{\pm}$ invariant solutions of $D=11$ supergravity \cite{dWNso7soln, englert} 
from the scalar expectation values of the SO$(7)^{\pm}$ invariant stationary 
points \cite{Warner83} of maximal SO$(8)$ gauged supergravity.  The possibility to perform
such explicit checks of all formulae against various non-trivial compactifications is a feature
that distinguishes our formalism from other approaches to generalised geometry.

From a four-dimensional point of view, the 56 vector fields include the full set of electric and magnetic vectors that can be gauged. In section \ref{sec:gvp}, we show that the generalised vielbein postulates determine exactly which of the vector fields are gauged in the $S^7$ compactification. In particular, we show that upon inserting the ans\"atze relevant for the $S^7$ compactification given in section \ref{sec:NLA}, the magnetic vector fields, which come from the reduction of the 3-form potential, drop out of the expressions. Moreover, the generalised vielbein postulates reduce to the E$_{7(7)}$ Cartan equation with SO$(8)$ gauge covariant derivative. The SO$(8)$ gauge fields are solely electric and arise from the graviphoton and the 6-form potential. More generally, for any compactification, the generalised vielbein postulates reduce to the four-dimensional Cartan equation with the appropriate gauge covariant derivative. Therefore, the generalised vielbein postulates provide an 
understanding of how the gauge vectors are selected from the 56 vector fields available. This goes some way towards 
establishing the origin of the embedding tensor in the higher-dimensional $D=11$ theory. 

We conclude in section \ref{sec:outlook} with a brief, general discussion of other compactifications that could lead to more general gaugings in four dimensions. Of particular interest are examples of compactifications where both electric and magnetic vectors are gauged. We also discuss the possibility of using our framework to provide an eleven-dimensional perspective on the recently discovered continuous family of SO$(8)$ gauged supergravities \cite{DIT}. 

In summary, the key results and issues raised in this paper are:
\begin{itemize}
\item Kaluza-Klein theory is developed for `non-geometric' vector fields.
\item Consistent non-linear ans\"atze are obtained for all fields, including dual fields.
\item All formulae can be tested against non-trivial compactifications
          of $D=11$ supergravity and the associated stationary points of $N=8$ 
          supergravity.
\item With 56 `electric' and `magnetic' vectors present in all $D=11$ relations,
          one can now study the higher-dimensional origins of the embedding tensor.
\item Preliminary evidence is presented that the $\omega$-deformed SO(8) gaugings
          of \cite{DIT} correspond to $\omega$-deformations of $D=11$ supergravity.
\end{itemize}

The conventions and index notations used in this paper are as in \cite{dWNsu8,GGN13}.

\section{Dual fields and E$_{7(7)}$ in $D=11$ supergravity} \label{sec:E711d}

In Ref.~\cite{GGN13}, the two generalised vielbeine previously known
from the literature \cite{dWNsu8,dWN13}, are completed to an E$_{7(7)}$
matrix in eleven dimensions by constructing two further generalised
vielbeine that are intimately related to the dualisation of the metric
and the three-form potential in eleven dimensions. While the significance 
of this construction from an eleven-dimensional point of view is clear 
in that it establishes the role of the E$_{7(7)}$ duality
group in the full $D=11$ theory, its importance from the practical point
of view of relating the four-dimensional maximal supergravity to $D=11$
supergravity is what will be addressed in this paper.  In particular,
the results of \cite{GGN13} give uplift ans\"atze for \emph{all} bosonic
degrees of freedom including a non-linear ansatz for the six-form
potential dual to the three-form potential. We will illustrate this in great detail
for the $S^7$ compactification of $D=11$ supergravity in the following sections, 
but from the generality of our results it should be clear that our construction 
furnishes similar information for other compactifications of $D=11$ supergravity.
Consequently, we will first summarise the general results in this section, without
reference to any specific compactification.

\subsection{Dual vector fields and generalised vielbeine} \label{sec:dual}

Working in the context of the SU$(8)$ invariant reformulation of $D=11$
supergravity one identifies certain SU$(8)$ objects starting from an analysis 
of the fermionic sector \cite{dWNsu8}.~\footnote{See also section 3.1 of
Ref.~\cite{GGN13} for a brief description of the SU$(8)$ invariant
reformulation.} In the bosonic sector the most prominent 
of these objects are the so-called
{\em generalised vielbeine}, which can be regarded as components of an
E$_{7(7)}$ matrix in eleven dimensions. The generalised vielbeine appear 
when one considers the supersymmetry transformation of those components 
of the elfbein, three-form potential and their dual fields which in a proper
reduction to four dimensions would give rise to vector fields. However, it is important
to keep in mind the main feature of the present analysis (and of \cite{dWNsu8}),
namely that {\em we retain the full coordinate dependence on all eleven 
coordinates throughout}. Therefore, we will not be dealing with a dimensional reduction 
in the strict sense of the word, but rather a 4+7 split and a subsequent reformulation
of the theory. Furthermore, the reformulated theory is on-shell equivalent to the 
original $D=11$ supergravity of \cite{CJS} {\em at all stages of the construction}.

Let us therefore first consider the spin-one sector of the theory. In the direct dimensional 
reduction of $D=11$ supergravity to four dimensions there appear only 28 vector 
fields, namely \cite{so(8)}
\begin{equation}\label{vec1} 
{B_{\mu}}^{m} \quad \mbox{and} \quad
B_{\mu m n } = A_{\mu m n } - B_{\mu}{}^{p} A_{pmn}.
\end{equation} 
The first seven of these are just the standard Kaluza-Klein vector fields in the decomposition 
of the elfbein displayed in \eqref{EMA}, while the second set of vectors originates
from the three-form field $A_{MNP}$. As explained in our previous work \cite{GGN13},
this set of vector fields is complemented by another set of vectors related to
the dual fields in eleven dimensions, {\it viz.}~\footnote{Note the slight change in
notation here compared with that used in \cite{GGN13}.  In particular,
here, we reserve the notation $B_{\mu m_1 \dots m_5}$ for the vector
whose supersymmetry transformation gives rise to the generalised
vielbein $e_{m_1 \ldots m_5 AB}$. A similar change of notation is made
for the fourth vector.  In addition, the arbitrary constant $\tilde{c}$
here is related to $c$ in \cite{GGN13} by $\tilde{c}=5! c/\sqrt{2}$.}
\begin{equation}
B_{\mu m_1 \dots m_5} = A_{\mu m_1 \dots m_5} - B_{\mu}{}^{p} A_{p m_1
\dots m_5} - \frac{\sqrt{2}}{4} \left( A_{\mu [m_1 m_2} - B_{\mu}{}^{p}
A_{p[m_1 m_2} \right) A_{m_3 m_4 m_5]}, \label{vec3}
\end{equation}
\begin{align}
& B_{\mu m_1 \dots m_7, n} = A_{\mu m_1 \dots m_7, n} + (3 \tilde{c}-1)
\left( A_{\mu [ m_1 \dots m_5} - B_{\mu}{}^{p} A_{p [m_1 \dots m_5}
\right) A_{m_6 m_7] n} \notag \\[3mm]
& \hspace{55mm}+ \tilde{c} A_{[ m_1 \dots m_6} \left( A_{|\mu|m_7] n} -
B_{\mu}{}^{p} A_{|p|m_7] n} \right) \notag \\[1mm]
 & \hspace{70mm} + \frac{\sqrt{2}}{12} \left( A_{\mu [m_1 m_2} -
B_{\mu}{}^{p} A_{p[m_1 m_2} \right) A_{m_3 m_4 m_5} A_{m_6 m_7] n}.
\label{vec4}
\end{align}
The vector components in \eqref{vec3} thus originate from the six-form $A_{(6)}$ which is 
the dual potential associated with the three-form potential $A_{(3)}$, whence it is clear that these fields are simultaneously defined only on-shell, as explained in \cite{GGN13}. The vector 
fields \eqref{vec4} are related to the dual gravity field in eleven dimensions, and  are
defined only up to a real constant $\tilde{c}$.  While the precise relation 
of $A_{\mu m_1 \dots m_7, n}$ to the $D=11$ fields is not known, the indeterminacy
encoded in the parameter $\tilde{c}$ can be traced back to the fact that
dual gravity does not give rise to scalar degrees of freedom in the 4+7 split.
We also note that the non-linear modifications in these equations which involve
the Kaluza-Klein vectors $B_\mu{}^m$ can be understood geometrically via the 
conversion of curved to flat indices, whereas the remaining non-linear modifications 
are required by the consistency of the supersymmetry variations, but have no direct 
explanation in terms of eleven-dimensional geometry.~\footnote{A geometrical explanation 
 might, however, follow from E$_{10}$ where the `vielbein' comprises not only the
 gravitational, but also the three- and six-form fields.}

Thus, in all we have identified 56 such vector fields in eleven dimensions, starting from 
the fields of $D=11$ supergravity and their duals. These make up part 
of the bosonic sector of our reformulation of $D=11$ supergravity in the framework
of the ``generalised geometry" introduced in \cite{GGN13}. However,
not all of these vector fields will correspond to independent propagating vectors in 
a given compactification of the $D=11$ theory. In particular, for compactifications
related to $N=8$ supergravity and deformations thereof,
we know that there can be at most 28 propagating spin-one degrees of freedom.
This is most easily seen in the $T^7$ reduction of \cite{so(8)}, where the seven `electric' 
vectors from $B_\mu{}^m$ (corresponding to the seven Killing vectors on $T^7$) combine
with 21 `magnetic' vectors from $B_{\mu mn}$ to give 28 abelian vector fields.  The other 28 vectors correspond to their four-dimensional duals such that
the eleven-dimensional duality relations reduce to the `twisted  self duality constraint'
of \cite{so(8)} in the reduction to four dimensions. For non-trivial compactifications 
of the theory the situation is, however,  much more complicated because of the 
appearance of {\em non-abelian} gauge interactions, for which the usual (abelian)
dualisation of vector fields does not work.

A judicious analysis of the supersymmetry 
transformations of these 56 vector fields \cite{dWNsu8,dWN13, GGN13} leads to the generalised vielbeine.
For the vector fields \eqref{vec1} the latter can be directly obtained from the 
$D=11$ theory, while the variation of $B_{\mu m_1\dots m_5}$ in \eqref{vec3} is 
determined from the variation of $A_{(6)}$ \cite{GGN13}.
The supersymmetry transformation of $B_{\mu m_1\dots m_7,n}$ 
is also given in Ref.~\cite{GGN13}, but it cannot be obtained from the $D=11$ theory. It can, however, be obtained by imposing consistency with the supersymmetry variations of the 
other vector fields. Somewhat lengthy computations show that \cite{GGN13}
\begin{align} \label{11dsusy1}
 \delta B_{\mu}{}^{m} &= \frac{\sqrt{2}}{8} \ e^{m}_{AB} \left[ 2
\sqrt{2} \overline{\varepsilon}^{A} \varphi_{\mu}^{B} +
\overline{\varepsilon}_{C} \gamma'_{\mu} \chi^{ABC} \right] \, +\,
\textup{h.c.}, \\[2mm] \label{11dsusy2}
  \delta B_{\mu m n} &= \frac{\sqrt{2}}{8} \ e_{mn AB} \left[ 2 \sqrt{2}
\overline{\varepsilon}^{A} \varphi_{\mu}^{B} +
\overline{\varepsilon}_{C} \gamma'_{\mu} \chi^{ABC} \right] \, +
\,\textup{h.c.}, \\[2mm] \label{11dsusy3}
   \delta B_{\mu m_1 \dots m_5} &= \frac{\sqrt{2}}{8} \ e_{m_1 \dots m_5
AB} \left[ 2 \sqrt{2} \overline{\varepsilon}^{A} \varphi_{\mu}^{B} +
\overline{\varepsilon}_{C} \gamma'_{\mu} \chi^{ABC} \right]  \, + \,
\textup{h.c.},\\[2mm]
    \delta B_{\mu m_1 \dots m_7, n} &= \frac{\sqrt{2}}{8} \ e_{m_1 \dots
m_7, n AB} \left[ 2 \sqrt{2} \overline{\varepsilon}^{A}
\varphi_{\mu}^{B} + \overline{\varepsilon}_{C} \gamma'_{\mu} \chi^{ABC}
\right] \, + \,  \textup{h.c.}, \label{11dsusy4}
\end{align}
where $\varphi_{\mu}^{A}$ and $\chi^{ABC}$ are the (chiral) fermions in the SU(8)
invariant reformulation, and where $\gamma'_\mu = e'_\mu{}^\alpha \gamma_\alpha$ 
with the Weyl rescaled vierbein $e'_\mu{}^\alpha$ from \eqref{EMA}.
The scalar coefficients in front of the fermionic bilinears make up the 
generalised vielbeine, and are given by  \cite{dWNsu8,dWN13,GGN13}
\begin{align}
e^{m}_{AB} &= i\Delta^{-1/2} \Gamma^m_{AB}, \label{gv1} \\[3mm]
e_{mn}{}_{AB} &= - \frac{\sqrt{2}}{12} i \Delta^{-1/2} \left(
\Gamma_{mn}{}_{AB} + 6 \sqrt{2} A_{mnp} \Gamma^{p}_{AB} \right),
\label{gv2} \\[3mm]
e_{m_1 \dots m_5}{}_{AB}&= \frac{1}{6!\sqrt{2}} i \Delta^{-1/2} \Bigg[
\Gamma_{m_1 \dots m_5}{}_{AB} + 60 \sqrt{2} A_{[m_1 m_2 m_3} \Gamma_{m_4
m_5]}{}_{AB} \notag \\
 & \hspace*{50mm}  - 6! \sqrt{2} \Big( A_{p m_1 \dots m_5} - 
\frac{\sqrt{2}}{4} A_{p[m_1 m_2} A_{m_3 m_4 m_5]} \Big) \Gamma^{p}_{AB}
\Bigg], \label{gv3} \\[2mm]
e_{m_1 \dots m_7, n}{}_{AB} &=- \frac{2}{9!} i \Delta^{-1/2} \Bigg[
(\Gamma_{m_1 \dots m_7} \Gamma_{n}{})_{AB} + 126 \sqrt{2}\ A_{n [m_1
m_2} \Gamma_{m_3 \dots m_7]}{}_{AB} \notag \\
 & \hspace*{37mm}  + 3\sqrt{2} \times 7! \Big( A_{n [ m_1 \dots m_5} +
\frac{\sqrt{2}}{4} A_{n[m_1 m_2} A_{m_3 m_4 m_5} \Big) \Gamma_{m_6
m_7]}{}_{AB} \notag \\[2mm]
  & \hspace*{42.5mm} + \frac{9!}{2} \Big(A_{n [ m_1 \dots m_5} +
\frac{\sqrt{2}}{12} A_{n[m_1 m_2} A_{m_3 m_4 m_5} \Big) A_{m_6 m_7] p }
\Gamma^{p}{}_{AB} \Bigg]. \label{gv4}
\end{align}
These objects carry SU(8) indices $A,B,\dots$ and are to be regarded as SU(8)
tensors in a specific gauge, as explained in \cite{dWNsu8}. As also shown there,
one thereby enlarges the original tangent space symmetry from SO(7) to a local
SU(8) symmetry that acts on the chiral fermions. Observe that the
new vielbein components, $e_{m_1 \dots m_5}{}_{AB}$ and
$e_{m_1 \dots m_7, n}{}_{AB}$, which originate from the variations of the dual vectors 
\eqref{vec3} and \eqref{vec4}, themselves depend on the dual six-form field
$A_{m_1\dots m_6}$, and hence again are only defined on-shell.

\subsection{Emergence of E$_{7(7)}$ structure}

One can make the relation of the above expressions to E$_{7(7)}$ more explicit by 
combining the matrix blocks into a single 56-bein {\em in eleven dimensions} \cite{GGN13}
\begin{equation}\label{56bein}
 \mathcal{V}(z)=\big(\cV^{\tM \tN}{}_{AB}(z) \,, \, \cV_{\tM \tN\, AB}(z)\big)
\end{equation}
by means of the identifications
\begin{align}
 \cV^{m8}{}_{AB} &= \frac{\sqrt{2}}{8} e^m_{AB}, \hspace{36mm} \cV_{mn\,
AB} = -\frac{3}{2} e_{mnAB}, \notag \\[2mm]
\cV^{mn}{}_{AB} &= -\frac{3}{2} \Delta \epsilon^{mnp_1\dots p_5}
e_{p_1\dots p_5 AB}, \hspace{10mm}
\cV_{m8\,AB} = \frac{9\sqrt{2}}{2} \Delta \epsilon^{n_1\dots n_7}
e_{n_1\dots n_7, m AB}. 
\label{Vdef}
\end{align}
This E$_{7(7)}$ vielbein is equivalent (see \cite{GGN13}) to the one considered in 
Ref.~\cite{hillmann} in the context of another proposal to realise an exceptional geometry. 
Note that complex conjugation acts by raising or lowering the SU(8) indices, {\it viz.}
\begin{equation}
\big( \cV^{\tM \tN}{}_{AB}\big)^* =   \cV^{\tM \tN\, AB}\;\; , \quad
\big( \cV_{\tM \tN}{}^{AB}\big)^* =   \cV_{\tM \tN\, AB},
\end{equation}
but leaves the position of the SL(8,$\mathbb{R}$) indices $\tM,\tN$ unaffected.
The 56-bein $\cV(z)$ as defined above is a coset element of
E$_{7(7)}$/SU$(8)$ written in terms of the decomposition of the $\bf 56$
of E$_{7(7)}$ under initially its SL$(8,\mathbb{R})$ and then
GL$(7,\mathbb{R})$ subgroups 
\begin{equation}
{\bf 56} \; \rightarrow \; {\bf 28} \oplus \overline{\bf 28} \;
\rightarrow \; {\bf 7} \oplus {\bf 21} \oplus \overline{\bf 21} \oplus
\overline{\bf 7}.
\end{equation}
The world indices $m,n,\dots$ labeling the seven-dimensional directions 
and originally transforming under seven-dimensional diffeomorphisms
thus become associated with the GL(7,$\mathbb{R}$) subgroup of E$_{7(7)}$.
In contrast to \cite{GGN13} we have adjusted the normalisation of the matrix
blocks in such a way that $\cV(z)$, as defined in \eqref{Vdef} satisfies the 
following identity 
\begin{equation}\label{sympl}
 \cV_{\tM \tN AB} \cV^{\tM \tN CD} - {\cV^{\tM \tN}}_{AB} {\cV_{\tM
\tN}}^{CD} = i \delta_{AB}^{CD}.
\end{equation}
This  is simply the statement that the inverse of an E$_{7(7)}$ matrix
is related to its complex conjugate; more specifically, \eqref{sympl} is a 
necessary condition expressing the fact that any E$_{7(7)}$ matrix 
automatically belongs to Sp(56,$\mathbb{R}$) \cite{so(8), dWNn8}. The
direct verification of \eqref{sympl} by substitution of \eqref{gv1}--\eqref{gv4} into 
\eqref{Vdef} is straightforward.

The vectors can be arranged into a similar object of the form
$({\mathcal{B}_{\mu}}^{\tM \tN}, \mathcal{B}_{\mu \, \tM \tN})$ such that
\begin{align}
  {\mathcal{B}_{\mu}}^{m 8} &= -\frac{1}{2} {B_{\mu}}^m, \hspace{36mm}
\mathcal{B}_{\mu\, mn} = 3\sqrt{2}\, B_{\mu mn}, \notag \\[2mm]
{\mathcal{B}_{\mu}}^{m n} &= 3\sqrt{2}\, \Delta \epsilon^{mnp_1\dots
p_5} B_{\mu p_1 \ldots p_5}, \hspace{10mm}
\mathcal{B}_{\mu\, m8} = -18\, \Delta \epsilon^{n_1\dots n_7} B_{\mu n_1
\dots n_7, m}. 
\label{Bdef}
\end{align}
Thereby the 28+28  vectors are combined into a $\bf{56}$ of E$_{7(7)}$.
In this language, the supersymmetry transformations
\eqref{11dsusy1}--\eqref{11dsusy4} can be compactly written as
\begin{align} \label{susycalB}
 \delta {\mathcal{B}_{\mu}}^{\tM \tN} &= -\frac{1}{2}
\cV^{\tM\tN}{}_{AB} \left[ 2 \sqrt{2} \overline{\varepsilon}^{A}
\varphi_{\mu}^{B} + \overline{\varepsilon}_{C} \gamma'_{\mu} \chi^{ABC}
\right] \,+ \, \textup{h.c.}, \notag \\[3mm]
 \delta \mathcal{B}_{\mu \, \tM \tN} &= -\frac{1}{2} \cV_{\tM\tN \, AB}
\left[ 2 \sqrt{2} \overline{\varepsilon}^{A} \varphi_{\mu}^{B} +
\overline{\varepsilon}_{C} \gamma'_{\mu} \chi^{ABC} \right] \, + \, \textup{h.c.}.
\end{align}
In \cite{GGN13} we have also shown that the matrix blocks making up the
56-bein \eqref{56bein} transform uniformly under local supersymmetry
\begin{equation}\label{susycalV}
\delta\cV^{\tM \tN}{}_{AB}(z) \,=\, -\sqrt{2} \Sigma_{ABCD} \cV^{\tM \tN\, CD}(z)
\; \, , \quad
\delta\cV_{\tM \tN \, AB}(z) \,=\, -\sqrt{2} \Sigma_{ABCD} \cV_{\tM \tN}{}^{CD}(z)
\end{equation}
with the complex self-dual tensor
\begin{equation}
\Sigma_{ABCD} \equiv \overline{\varepsilon}_{[A} \chi_{BCD]} \, +    \,
\frac1{24} \epsilon_{ABCDEFGH} \overline{\varepsilon}^{E} \chi^{FGH} 
\end{equation}
where we have discarded a local SU(8) rotation that also acts uniformly on all components.
In the form \eqref{susycalB} and \eqref{susycalV} the supersymmetry variations of 
compactified maximal supergravity can be read off directly from the $D=11$ formulae.

\subsection{Generalised vielbein postulates} \label{subsec:gvp}

The generalised vielbeine satisfy certain differential constraints derived in
\cite{dWNsu8, GGN13}. These constraints are generalisations of the vielbein 
postulate in Riemannian geometry which establishes the relation between 
affine and spin connections. For our subsequent analysis, here we only need 
their components along the four space-time directions, which are \cite{dWNsu8, GGN13}
\begin{align} \label{cDB1}
 & \cD_{\mu} e^{m}_{AB} + \frac{1}{2} \partial_{n} B_{\mu}{}^{n} e^{m}_{AB} + \partial_{n} B_{\mu}{}^{m} e^{n}_{AB} + \cQ_{\mu}^{C}{}_{[A} e^{m}_{B]C} + \cP_{\mu ABCD} e^{mCD} = 0, \\[3mm]
\label{cDB2}
&\cD_\mu e_{mnAB} + \frac{1}{2} \partial_{p} B_{\mu}{}^{p} e_{mnAB} + 2 \partial_{[m} 
B_{|\mu|}{}^{p} e_{n]pAB} + 3 \partial_{[m} B_{|\mu|np]} e^{p}_{AB} \notag \\[2pt]
&\hspace{80mm}+ \cQ_{\mu}^{C}{}_{[A} e_{mnB]C} + \cP_{\mu ABCD} e_{mn}{}^{CD} = 0, \\[3mm]
&\cD_\mu e_{m_1 \dots m_5 AB} + \frac{1}{2} \partial_{p} B_{\mu}{}^{p} e_{m_1 \dots m_5 AB} - 5 \partial_{[m_1} B_{|\mu|}{}^{p} e_{m_2 \dots m_5]pAB} + 
\frac{3}{\sqrt{2}} \partial_{[m_1} B_{|\mu|m_2 m_3} e_{m_4 m_5]AB} \notag \\[3pt] \label{cDB3}
& \hspace{40mm}- 6 \partial_{[m_1} B_{|\mu| m_2 \dots m_5 p]} e^{p}_{AB} + \cQ_{\mu}^{C}{}_{[A} e_{m_1 \dots m_5 B]C} + \cP_{\mu ABCD} e_{m_1 \dots m_5}{}^{CD}=0, \\[10pt]
& \cD_\mu e_{m_1 \dots m_7, n AB} - \frac{1}{2} \partial_{p} B_{\mu}{}^{p} e_{m_1 \dots m_7, n AB} - \partial_{n} B_{\mu}{}^{p} e_{m_1 \dots m_7, p AB} + 5 \partial_{[m_1} B_{|\mu| m_2 m_3} e_{m_4 \dots m_7] n AB}  \notag \\[2mm] 
& \hspace{30mm} - 2 \partial_{[m_1} B_{|\mu| m_2 \dots m_6} e_{m_7] n AB}+ \cQ_{\mu}^{C}{}_{[A} e_{m_1 \dots m_7, n B]C} + \cP_{\mu ABCD} e_{m_1 \dots m_7, n}{}^{CD} =0, \label{cDB4}
\end{align}
where~\footnote{Below we will work with a slightly modified operator $\cD_\mu$ adapted to the $S^7$ compactification, cf. \eqref{cDmu}.} 
\begin{equation}\label{cD0}
\cD_\mu \equiv \partial_\mu - B_\mu{}^m \partial_m.
\end{equation}
In comparison with  equation (88) of \cite{GGN13} 
the generalised vielbein postulate for $e_{m_1 \dots m_7, n AB}$ has been 
simplified by using the Schouten identity over eight indices. The connection coefficients 
$Q_\mu$ and $P_\mu$ appearing in these equations are valued in the  E$_{7(7)}$ 
Lie algebra, and are related to the connections and four-form field strengths
of $D=11$ supergravity as follows \cite{dWNsu8}
\begin{align}\label{GVP0}
 \cQ_{\mu}^{A}{}_{B} &= - \textstyle{\frac{1}{2}} \left[ {e^m}_a \partial_{m} B_{\mu}{}^{n} e_{n b} - ({e^p}_{a} \cD_{\mu} e_{p\, b}) \right] \Gamma^{ab}_{AB} 
 - \textstyle{\frac{\sqrt{2}}{12}} \Delta^{-1/2} {e'_{\mu}}{}^{\alpha} \left( F_{\alpha abc} \Gamma^{abc}_{AB} - \eta_{\alpha \beta \gamma \delta} F^{\beta \gamma \delta a} \Gamma_{a AB} \right), \\[3mm]
 \cP_{\mu ABCD}& = \textstyle{\frac{3}{4}} \left[ {e^m}_a \partial_{m} B_{\mu}{}^{n} e_{n b} - ({e^p}_{a} \cD_{\mu} e_{p\, b}) \right] \Gamma^{a}_{[AB} \Gamma^{b}_{CD]}
 - \textstyle{\frac{\sqrt{2}}{8}} \Delta^{-1/2} {e'_{\mu}}{}^{\alpha} F_{abc \alpha} \Gamma^{a}_{[AB} \Gamma^{bc}_{CD]} \notag \\[2mm]
& \hspace{75mm} - \textstyle{\frac{\sqrt{2}}{48}} \Delta^{-1/2} e'_{\mu \, \alpha} \eta^{\alpha \beta \gamma \delta} F_{a \beta \gamma \delta}{\Gamma_{b}}_{[AB} \Gamma^{ab}_{CD]}.
\label{GVP1}
\end{align}
Below we will use them in a slightly modified form, again adapted to the $S^7$ compactification.

\section{Non-linear ans\"atze for maximal supergravity on $S^7$}  \label{sec:NLA}

We will now illustrate the usefulness of the results in the foregoing section
by specialising to the $S^7$ compactification of $D=11$ supergravity \cite{freundrubin, duffpope}, where
our formalism furnishes numerous new insights and results, most notably with 
regard to the dual fields of $D=11$ supergravity. To this aim, we will present a detailed
analysis of the non-linear ans\"atze and the generalised vielbein postulate
in the context of the $S^7$ compactification.

\subsection{Compactification on $S^7$}

To begin with, and for the reader's convenience, here we collect some relevant 
(and well-known) formulae, see e.g. \cite{duffpope,GGN}, related to the 
seven-sphere. Denoting the $S^7$ background covariant 
derivative by $\Do_m$, we recall that $S^7$ admits eight Killing 
spinors $\eta^I(y)$ obeying
\begin{equation}
 \left(\Do_{m} + \textstyle{\frac{1}{2}} i m_7 \eo_{m}{}^{a} \Gamma_{a}
\right) \eta^{I} =0,
\end{equation} 
where $I,J,...= 1,...,8$, and $\eo_m{}^a$ is the siebenbein on the round $S^7$.
Written out in components, the Killing spinors $\eta^I_A$
are orthonormal matrices, that is
\begin{equation}\label{orthoKS}
\eta^A_I \eta^J_A = \delta^I_J,  \;\;\quad \eta^A_I \eta^I_B = \delta^A_B.
\end{equation}
The 28 Killing vectors $K^{mIJ}(y)$ and their derivatives $K_{mn}{}^{IJ}(y)$ can 
then be represented as bilinears in terms of Killing spinors, {\it viz.}
\begin{equation}
 K^{m IJ} = i \eo^{ma} \, \overline{\eta}^{I} \Gamma_{a} \eta^{J}\, ,
\qquad  K_{m n}{}^{IJ} = \eo_{m}{}^{a} \eo_{n}{}^{b} \, \overline{\eta}^{I}
\Gamma_{ab} \eta^{J}\, ; \label{Kdefs} 
\end{equation}
clearly,
\begin{equation}
 \Do_{n} K_{m}^{IJ} = m_7 K_{m n}^{I J}     \label{K1deriv}
 \quad \Rightarrow \quad    \Do_{m} K_{n}^{IJ} +  \Do_{n} K_{m}^{IJ} = 0 \, .
 \end{equation}
Hence,  $\Do_{n} K^{n IJ} =0$.   Observe the different `canonical' positions  of the 
world indices on $K^m$ and $K_{mn}$, and the fact that both of these are always 
and by definition related to the `flat' objects by means of the $S^7$ background 
siebenbein and its inverse. The vector fields $K^{IJ}\equiv K^{mIJ} \Do_m$ 
generate the SO(8) isometry group of the seven-sphere via the Lie bracket 
\begin{equation}\label{KVcomm1}
\left[ K^{IJ}, K^{KL} \right] 
= - 8 m_7 \delta^{[I|[K} K^{L]|J]}
\end{equation}
or equivalently
\begin{gather}\label{KVcomm2}
 K^{n IJ} \Do_{n} K^{m KL} - K^{n KL} \Do_{n} K^{m IJ} = - 8 m_7 \delta^{[I|[K} K^{m L]|J]}. 
\end{gather}
However, in standard Kaluza-Klein geometry there is no corresponding
interpretation for the tensor fields $K_{mn}{}^{IJ}$ (nor for \eqref{Km5} below).

\subsection{The non-linear ans\"atze}

The non-linear ans\"atze for maximal supergravity are obtained by
comparing the supersymmetry transformations \eqref{susycalB}, with the
analogous supersymmetry transformations of the vectors in
$D=4$ maximal supergravity \cite{dWNn8, dWSTmax4}:
\begin{align}
 \delta {A_{\mu}}^{IJ} &= - \textstyle{\frac{1}{2}}
({{u_{ij}}^{IJ}}+v_{ijIJ}) \left[ 2 \sqrt{2} \overline{\varepsilon}^{i}
\varphi_{\mu}^{j} + \overline{\varepsilon}_{k} \gamma'_{\mu} \chi^{ijk}
\right]
 \, + \, \textup{h.c.}, \label{susyev} \\[1mm]
 \delta A_{\mu IJ} &= - \textstyle{\frac{1}{2}}
i({{u_{ij}}^{IJ}}-v_{ijIJ}) \left[ 2 \sqrt{2} \overline{\varepsilon}^{i}
\varphi_{\mu}^{j} + \overline{\varepsilon}_{k} \gamma'_{\mu} \chi^{ijk}
\right]
 \, + \, \textup{h.c.} \,.  \label{susymv}
\end{align}
We will refer to the 28 + 28 vector fields ${A_{\mu}}^{IJ}$ and $A_{\mu IJ}$  as `electric'
and `magnetic' vectors, respectively.

In order to relate the $D=4$ vectors to the vector fields identified in the previous
section,  we now need to choose appropriate Kaluza-Klein ans\"atze 
for all vectors \eqref{vec1}, \eqref{vec3} and \eqref{vec4}. For the Kaluza-Klein vector  
$B_\mu{}^m$ this ansatz is well-known, as we explained in the introduction;
choosing appropriate normalisations, we have
\begin{equation}
 {B_{\mu}}^{m}(x,y) = - \frac{\sqrt{2}}{4} K^{mIJ}(y)
{A_{\mu}}^{IJ}(x), \label{vec14d} 
\end{equation}
where $K^{mIJ}$ are the 28 Killing vectors defined in equation \eqref{Kdefs}.
However, for the remaining vector fields, and in particular for those 
arising from the dual fields in higher dimensions, there are no such
ans\"atze available from general Kaluza-Klein theory, and therefore 
we have to proceed in a different manner. In fact, for the `non-geometrical' 
vector fields $B_{\mu mn}(x,y)$ the appropriate ansatz was already 
found in \cite{dWN13}; it reads~\footnote{\label{foot:dis} Note that the difference
between the coefficient in the ansatz for $B_{\mu mn}$ here and in Ref.
\cite{dWN13} is due to differing conventions for $F= \d A.$ In
\cite{dWN13}, this constant was fixed based on the tests of the
non-linear flux ansatz in \cite{GGN} where $F= 4 \partial A.$ However,
here, as in much of the related literature, we use the convention $ F=
4! \partial A.$}
\begin{equation}
 B_{\mu m n}(x,y) = \frac{1}{24} {K_{mn}}^{IJ}(y) A_{\mu IJ}(x).
\label{vec24d}
\end{equation}
We stress that the `canonical' position of the world indices
as defined in Ref.~\cite{dWN13} is in accord with the position of indices on $B_\mu{}^m$ and 
$B_{\mu mn}$. We emphasise again that, unlike for the standard Kaluza-Klein 
vector, there is {\it a priori} no geometric argument to fix the ansatz for $B_{\mu mn}$.
Furthermore, the normalisation had to be determined in \cite{GGN} by comparison with
the $D=4$ theory.

Adopting the Kaluza-Klein ans\"atze \eqref{vec14d} and \eqref{vec24d} for $B_\mu{}^m$ 
and $B_{\mu mn}$, respectively, and comparing the eleven-dimensional supersymmetry transformations \eqref{11dsusy1} and \eqref{11dsusy2} with the respective
four-dimensional transformations \eqref{susyev} and \eqref{susymv} gives
a relation between the generalised vielbeine $e^{m}_{AB}$ and $e_{mn}{}_{AB}$ 
and the four-dimensional scalars ${{u_{ij}}^{IJ}}$ and $v_{ijIJ}$. More precisely, 
the identification between 11-dimensional and 4-dimensional SU(8) indices is made 
by means of the orthonormal Killing spinors on the round sphere $\eta^{i}_{A}$, which
convert `curved' SU(8) indices $A,B, \dots$ (appropriate to the $D=11$ theory)
into `flat' SU(8) indices $i,j,k,\dots$ (appropriate to maximal $D=4$ gauged supergravity)
and {\it vice versa} in the terminology of \cite{dWNconsis}. Hence,
\begin{equation}\label{SU8cf}
X_{ijk\cdots} = \eta_i^A \eta_j^B \eta_k^C \cdots  X_{ABC\cdots} 
\quad \Leftrightarrow\quad
X_{ABC\cdots} = \eta^i_A \eta^j_B \eta^k_C \cdots  X_{ijk\cdots} 
\end{equation}
for any SU(8) tensor by orthonormality of the Killing spinors \eqref{orthoKS}.
Accordingly, we define
\begin{eqnarray}
e^{m}_{ij} (x,y) &\equiv& e^{m}_{A B} (x,y) {\eta_{i}}^{A}(y) {\eta_{j}}^{B}(y)\nonumber\\[2mm] 
e_{mn\,ij} (x,y) &\equiv& e_{mn\,A B} (x,y) {\eta_{i}}^{A}(y) {\eta_{j}}^{B}(y).
\end{eqnarray}
The non-linear ans\"atze derived in previous work are then given by
\begin{align}
 e^{m}_{ij}(x, y) &=  K^{mIJ}(y) \big[{{u_{ij}}^{IJ}}+v_{ijIJ}\big](x),
\label{gv14d} \\[2mm]
 e_{mn}{}_{ij}(x, y) &= - \frac{\sqrt{2}}{12} i {K_{mn}}^{IJ}(y)
\big[{{u_{ij}}^{IJ}}-v_{ijIJ}\big](x). \label{gv24d}
\end{align}

The ans\"atze for the remaining vectors originating from the dual $D=11$ fields
are more tricky and, in fact, can only be arrived  at by imposing consistency
of the relevant supersymmetry variations.
More specifically, our analysis implies the identifications 
\begin{align}
B_{\mu m_1 \dots m_5}(x,y) &= - \frac{1}{4\cdot 6!} \left({K_{m_1 \dots
m_5}}^{IJ} - 6\cdot 6! \, \zetao_{m_1 \dots m_5 p} K^{p IJ} \right)(y)\
{A_{\mu}}^{IJ}(x), \label{vec34d} \\[2mm]
B_{\mu m_1 \dots m_7, n}(x,y) &= - \frac{1}{9!\sqrt{2}} \left(\etao_{m_1
\dots m_7} {K_{n}}^{IJ} + 6\cdot 7! \, \zetao_{[m_1 \dots m_6}
{K_{m_7]n}}^{IJ} \right)(y)\ A_{\mu IJ}(x), \label{vec44d}
\end{align}
where
\begin{equation}\label{Km5}
{K_{m_1 \dots m_5}}^{IJ} = i \eo_{m_1}{}^{a_1} \cdots \eo_{m_5}{}^{a_5} \
\bar{\eta}^I \Gamma_{a_1 \dots a_5} \eta^J
\end{equation}
is again a bilinear in the Killing spinors.
The quantity $\zetao_{m_1 \ldots m_6}(y)$ is defined such that
\begin{equation}
 7! \, \partial_{[m_1} \zetao_{m_2 \ldots m_7]} = m_7 \, \etao_{m_1
\ldots m_7}, \label{7volpot}
\end{equation}
and is thus to be regarded as a potential for the volume form on the round seven-sphere. 

While the first terms on the right hand side of the ans\"atze \eqref{vec34d} and \eqref{vec44d} are 
not completely unexpected, the crucial new feature in comparison with formulae
\eqref{vec14d} and \eqref{vec24d} is the presence of the non-globally defined field 
$\zetao_{m_1 \ldots m_6}$. One way of understanding its presence in the above
ans\"atze is to observe that the components of the field strength
$F_{(4)}$ along the four dimensions $F_{\mu \nu \rho \sigma}$ is
non-zero and, for maximally symmetric solutions is proportional
to the volume form in four dimensions \cite{freundrubin}.  Equivalently,
this non-zero component of the three-form potential $A_{\mu \nu \rho}$
can be viewed as a non-zero component of its six-form dual along the
seven-dimensional directions \cite{GGN13}, namely
\begin{equation}\label{Azeta}
 A_{m_1 \ldots m_6} \,\sim\,  \zetao_{m_1 \ldots m_6}.
\end{equation}
Thus, even for the $S^7$ solution \cite{duffpope} where the
scalar expectation values are essentially trivial, one would have to
obtain such a non-zero value for $A_{m_1 \ldots m_6}$ from a non-linear
ansatz.  Indeed, the coefficients in ans\"atze \eqref{vec34d} and
\eqref{vec44d} have been fixed by requiring consistency with the $S^7$
solution. In fact, the `vacuum expectation value' of the six-form
field will be non-vanishing for {\em any} non-trivial compactification
(that is, other than the $T^7$ reduction of \cite{so(8)}) of $D=11$ supergravity.

As before, comparing the four-dimensional supersymmetry transformations
\eqref{susyev} and \eqref{susymv} with their higher-dimensional
analogues \eqref{11dsusy3} and \eqref{11dsusy4} gives
\begin{align}
 e_{m_1 \dots m_5}{}_{ij}(x, y) &= \frac{1}{6!\sqrt{2}} \left({K_{m_1
\dots m_5}}^{IJ} - 6\cdot 6! \, \zetao_{m_1 \dots m_5 p} K^{p IJ}
\right)(y) \big[{{u_{ij}}^{IJ}}+v_{ijIJ}\big](x), \label{gv34d} \\[3mm]
 e_{m_1 \dots m_7, n}{}_{ij}(x, y) &= \frac{2}{9!}i \left(\etao_{m_1
\dots m_7} {K_{n}}^{IJ} + 6\cdot 7! \, \zetao_{[m_1 \dots m_6}
{K_{m_7]n}}^{IJ} \right)(y) \big[{{u_{ij}}^{IJ}}-v_{ijIJ}\big](x), \label{gv44d}
\end{align}
where we have again converted `curved' to `flat' SU(8) indices by
means of relations \eqref{SU8cf}.

The fact that we now have two expressions for the generalised vielbeine,
one in terms of eleven-dimensional fields, equations
\eqref{gv1}--\eqref{gv4}, and one in terms of four-dimensional scalars,
equations \eqref{gv14d}, \eqref{gv24d}, \eqref{gv34d} and \eqref{gv44d},
allows us to derive non-linear ans\"atze for internal fields.  The
non-linear ansatz for the metric \cite{dWNW} is found by considering the
expression
\begin{equation}
 {e^{m}}_{AB} e^{n AB},
\end{equation}
which gives \cite{dWNW}
\begin{equation}
\Delta^{-1} g^{mn}(x,y) = \frac{1}{8} K^{m}{}^{IJ} K^{n}{}^{KL}(y)
\Big[ (u^{ij}{}_{IJ} + v^{ijIJ}) (u_{ij}{}^{KL} + v_{ijKL}) \Big] (x) \, .
\label{metansatz}
\end{equation}
Similarly, the non-linear flux ansatz \cite{dWN13} is found by
considering the expression
\begin{equation}
 {e_{mn}}^{AB} {e^{p}}_{AB},
\end{equation}
which gives~\footnote{See footnote \ref{foot:dis} for an
explanation of the extra factor of 6 between this expression and
the non-linear flux ansatz in \cite{dWN13,GGN}.}
\begin{equation}
A_{mnp}(x,y) =- \frac{\sqrt{2}}{96}\ i \Delta\, g_{pq}(x,y)\, K_{mn}{}^{IJ}
K^{q}{}^{KL}(y)
\Big[( u^{ij}{}_{IJ} - v^{ij IJ}) ( u_{ij}{}^{KL} + v_{ij KL} )\Big](x)\, ,
\label{potans}
\end{equation}
where one uses the metric ansatz \eqref{metansatz} to compute $\Delta
g_{pq}$ in the equation above.  Now, considering
\begin{equation}
 e_{m_1 \ldots m_5 AB} e^{n AB}
\end{equation}
gives a non-linear ansatz for the internal six-form components.  It is
simple to show that by equating the contraction of the two vielbeine
using definitions \eqref{gv1} and \eqref{gv3} on one side and
definitions \eqref{gv14d} and \eqref{gv34d} on the other gives
\begin{align}
&A_{n m_1 \dots m_5} -  \frac{\sqrt{2}}{4} A_{n[m_1 m_2} A_{m_3 m_4
m_5]} \notag \\[4mm]
&\hspace{8mm}=-\frac{\sqrt{2}}{16\cdot 6!}\, \Delta\, g_{np}\,
\left({K_{m_1 \dots m_5}}^{IJ} - 6\cdot 6! \, \zetao_{m_1 \dots m_5 q}
K^{q IJ} \right) K^{p KL} \Big[({{u_{ij}}^{IJ}}+v_{ijIJ}) (
u^{ij}{}_{KL} + v^{ij KL})\Big],
\label{6formans}
\end{align}
where the internal metric and three-form potential components are
derived using ans\"atze \eqref{metansatz} and \eqref{potans}.  We note
that the complete antisymmetry of $A_{m_1\dots m_6}$ in all six
indices is not manifest from this expression; our explicit tests for the
SO$(7)^\pm$ invariant solutions (see section \ref{sec:test}) show, however, that this consistency requirement
is met in non-trivial examples.

An ansatz for the six-form potential can also be obtained by considering the fourth generalised vielbeine $e_{m_1 \ldots m_7, n AB}$.  The relation of this ansatz to the ansatz \eqref{6formans} above will not be obvious, but clearly the two ans\"atze must be equivalent.
This completes the set of uplift ans\"atze for all bosonic degrees of
freedom from maximal gauged supergravity to eleven dimensions.

\section{Generalised vielbein postulates and the $S^7$ compactification} \label{sec:gvp}

The generalised vielbein postulate for $e^{m}_{AB}$ plays an important role in 
establishing the consistency of the $S^7$ reduction of $D=11$ 
supergravity \cite{dWNconsis, NP}. In particular, in \cite{dWNconsis} it is shown that upon 
the $S^7$ compactification, the $d=4$ generalised vielbein postulate reduces to 
the E$_{7(7)}$ Cartan equation of gauged maximal supergravity, to wit
\begin{equation}\label{CM}
\cV^{-1}(x) \big( \partial_\mu \, -\, g A_\mu^{IJ}(x) X^{IJ}\big) \cV(x) 
= Q_\mu(x) + P_\mu (x),
\end{equation}
where $X^{IJ}$ generate the compact SO(8) subgroup inside SL(8,$\mathbb{R})\subset$
E$_{7(7)}$, and $g$ is the gauge coupling constant. In this section, then, 
we explore the generalised vielbein postulates 
in a more general context than \cite{dWNconsis,NP} by investigating the full set
of relations \eqref{cDB1}--\eqref{cDB4} for the extra vielbein components \eqref{gv1}--\eqref{gv4}, hence taking into account the full set of 56 vector fields 
identified in section \ref{sec:dual}. The presence of both electric {\em and} magnetic
vectors in these relations indicates that our construction should eventually allow one
to derive more general gaugings of $N=8$ supergravity from compactification, and 
thereby to understand how the embedding tensor emerges from the $D=11$ theory 
upon different compactifications. However, here we will concentrate on the
$S^7$ compactification, as this case already by itself provides a wealth of new 
insights, in particular concerning the role of dual vector fields in non-abelian
gaugings. We will briefly return to the more general case in the final section, 
postponing a detailed discussion to later work.

One issue that we will specifically address and resolve in this section is the following:
The fact that not all of the 56 vector fields can correspond to independent propagating 
degrees of freedom, and the generic emergence of {\em non-abelian} gauge 
interactions for non-trivial compactifications (for which the standard abelian dualisation
linking electric and magnetic vectors no longer works), immediately raises the question of
how the theory can dispose of the unwanted vectors and thereby ensure the 
consistency of compactified theory. Here we will establish consistency of the equations 
for the $S^7$ compactification by explicitly showing how the magnetic vector fields 
drop out of the generalised  vielbein postulates, leaving only electric gaugings. 

For the $S^7$ compactification the reduction ansatz of the vector fields and the 
generalised vielbeine are given by equations \eqref{vec14d}, \eqref{vec24d}, \eqref{vec34d}, \eqref{vec44d}, and equations \eqref{gv14d}, \eqref{gv24d}, \eqref{gv34d}, \eqref{gv44d}, respectively. To adapt to the $S^7$ compactification we introduce a minor modification
by replacing \eqref{cD0} by
\begin{equation}\label{cDmu}
 \cD_\mu = \partial_{\mu} - B_{\mu}{}^{m} \Do_{m},
\end{equation}
and the connections \eqref{GVP0} and \eqref{GVP1} by
\begin{align}
 \cQ_{\mu}^{A}{}_{B} &= - \textstyle{\frac{1}{2}} \left[ {e^m}_a \Do_{m} B_{\mu}{}^{n} e_{n b} - ({e^p}_{a} \cD_{\mu} e_{p\, b}) \right] \Gamma^{ab}_{AB} 
 - \textstyle{\frac{\sqrt{2}}{12}} \Delta^{-1/2} {e'_{\mu}}{}^{\alpha} \left( F_{\alpha abc} \Gamma^{abc}_{AB} - \eta_{\alpha \beta \gamma \delta} F^{\beta \gamma \delta a} \Gamma_{a AB} \right), \\[3mm]
 \cP_{\mu ABCD}& = \textstyle{\frac{3}{4}} \left[ {e^m}_a \Do_{m} B_{\mu}{}^{n} e_{n b} - ({e^p}_{a} \cD_{\mu} e_{p\, b}) \right] \Gamma^{a}_{[AB} \Gamma^{b}_{CD]}
 - \textstyle{\frac{\sqrt{2}}{8}} \Delta^{-1/2} {e'_{\mu}}{}^{\alpha} F_{abc \alpha} \Gamma^{a}_{[AB} \Gamma^{bc}_{CD]} \notag \\[2mm]
& \hspace{75mm} - \textstyle{\frac{\sqrt{2}}{48}} \Delta^{-1/2} e'_{\mu \, \alpha} \eta^{\alpha \beta \gamma \delta} F_{a \beta \gamma \delta}{\Gamma_{b}}_{[AB} \Gamma^{ab}_{CD]},
\end{align}
that is, replacing the partial derivative $\partial_m$ by the $S^7$ background covariant 
derivative $\Do_m$ everywhere. Likewise this replacement is to be made
everywhere in the vielbein postulate equations \eqref{cDB1}--\eqref{cDB4}.

As before (cf. \eqref{SU8cf}) we have to convert SU(8) indices in order to relate
the connection coefficients above to their four-dimensional counterparts for the
$S^7$ compactification. This change of basis is covariant for all fields, with the 
exception of \cite{dWNconsis}
\begin{equation}\label{QAmu}
 \cQ_{\mu}^{i}{}_{j} = \eta^{i}_{A} \eta^{B}_{j} \left( \cQ_{\mu}^{A}{}_{B} - i \textstyle{\frac{\sqrt{2}}{4}} m_7 A_{\mu}{}^{KL} K^{n KL} \eo_{n}{}^{a} \Gamma_{a A B} \right).
\end{equation}
Using the Killing spinor equation and the equation above, 
\begin{align} 
 \cD_\mu e^{m}_{AB} + \cQ_{\mu}^{C}{}_{[A} e^{m}_{B]C} &= \left(\partial_{\mu} - B_{\mu}{}^{m} \Do_{m}\right) \left[\eta_{A}^{i}(y) \eta^{j}_{B}(y) e^{m}_{i j}(x, y)\right] + \cQ_{\mu}^{C}{}_{[A} e^{m}_{B]C}, \notag \\[3pt]
 &= \eta_{A}^{i} \eta^{j}_{B} \cD_\mu e^{m}_{ij} + i m_7 B_{\mu}{}^{m} \eo_{m}{}^{a} \Gamma^{a}{}_{C[A} e^{m}_{B]C} + \cQ_{\mu}^{C}{}_{[A} e^{m}_{B]C}, \notag \\
 &= \eta_{A}^{i} \eta^{j}_{B} \left( \cD_\mu e^{m}_{ij} + \cQ_{\mu}^{k}{}_{[i} e^{m}_{j]k} \right). \label{Desu8trans}
\end{align}
Analogous relations hold for the other generalised vielbeine:
\begin{align} 
 \cD_\mu e_{mn AB} + \cQ_{\mu}^{C}{}_{[A} e_{mn B]C} &= \eta_{A}^{i} \eta^{j}_{B} \left( \cD_\mu e_{m n ij} + \cQ_{\mu}^{k}{}_{[i} e_{mn j]k} \right), \label{Desu8trans2} \\
  \cD_\mu e_{m_1 \dots m_5 AB} + \cQ_{\mu}^{C}{}_{[A} e_{m_1 \dots m_5 B]C} &= \eta_{A}^{i} \eta^{j}_{B} \left( \cD_\mu e_{m_1 \dots m_5 ij} + \cQ_{\mu}^{k}{}_{[i} e_{m_1 \dots m_5 j]k} \right), \label{Desu8trans3} \\
   \cD_\mu e_{m_1 \dots m_7, n AB} + \cQ_{\mu}^{C}{}_{[A} e_{m_1 \dots m_7, n B]C} &= \eta_{A}^{i} \eta^{j}_{B} \left( \cD_\mu e_{m_1 \dots m_7, n ij} + \cQ_{\mu}^{k}{}_{[i} e_{m_1 \dots m_7, n j]k} \right). \label{Desu8trans4}
\end{align}
The non-covariant term in \eqref{QAmu} thus ensures that we can freely convert between
`curved' and `flat' SU(8) indices in all relations.

Let us first consider the generalised vielbein postulate for $e^{m}_{AB},$ which is already analysed in \cite{dWNconsis}. The supersymmetry transformation of the graviphoton $B_{\mu}{}^{m}$ gives rise to the generalised vielbein $e^{m}_{AB},$ equation \eqref{11dsusy1}. In Kaluza-Klein theory, the exact ansatz relating the graviphoton to the four-dimensional vector 
field is given by the the Killing vectors of the internal space, \eqref{vec14d}. 
As we have already mentioned, this ansatz, via equations \eqref{susyev} and \eqref{11dsusy1}, also furnishes an ansatz \eqref{gv14d} for the generalised vielbein $e^{m}_{AB}$. 
The emergence of the SO(8) covariantisation of the four-dimensional derivative is then easily seen to be a consequence, in accordance with general Kaluza-Klein theory, 
of the appearance of the commutator of two Killing vector fields in
\begin{equation}
(\partial_\mu - B_\mu{}^n\Do_n) e^m_{ij} + \Do_n B_\mu{}^m e^n_{ij}
\end{equation}
and the fact that $\Do_m B_\mu{}^m = 0$ for any Killing vector.
More precisely, plugging these ans\"atze into the generalised vielbein 
postulate \eqref{cDB1} and using equation \eqref{Desu8trans}, the latter reduces to
\begin{align}
 \partial_{\mu} e^{m}_{ij} & - \frac{\sqrt{2}}{8} \Do_{n} K^{n IJ} {A_{\mu}}^{IJ} e^{m}_{ij} + \frac{\sqrt{2}}{4} \left(K^{n IJ} \Do_{n} K^{m KL} - K^{n KL} \Do_{n} K^{m IJ} \right) {A_{\mu}}^{IJ} 
 \big[{{u_{ij}}^{KL}}+v_{ijKL}\big] \notag \\[4mm]
  &\hspace{70mm} + \, \cQ_{\mu}^{k}{}_{[i} e^{m}_{j]k} \,+\, \cP_{\mu ijkl} e^{mkl} = 0. \label{gvp1}
\end{align}
Using equation \eqref{KVcomm2}, the first three terms in the generalised vielbein postulate \eqref{gvp1} reduce to
\begin{equation}
K^{m IJ} \left( \delta^{I}_{K} \partial_{\mu} - 2 \sqrt{2} {A_{\mu}}^{IK} \right) 
\big[{{u_{ij}}^{K J}}+v_{ijKJ}\big] = 0,
\end{equation}
which is a contraction of the SO(8) gauge covariant derivative on the scalar fields. Denoting
\begin{equation}
 {w^{+}}_{i j}{}^{IJ} = u_{ij}{}^{IJ} + v_{ijIJ}, \qquad 
  {w^{-}}_{i j}{}^{IJ} = i \big[u_{ij}{}^{IJ} - v_{ijIJ}\big]
\end{equation}
and  by ${w^{+}}^{i j}{}_{IJ}$ and ${w^{-}}^{i j}{}_{IJ}$ their complex conjugates, respectively, 
the generalised vielbein postulate gives
\begin{equation}
K^{m IJ} \left( D^{\textup{SO}(8)}_{\mu} {w^{+}}_{i j}{}^{IJ} + \cQ_{\mu}^{k}{}_{[i} {w^{+}}_{j]k}{}^{IJ} + \cP_{\mu ijkl} {w^{+}}^{kl}{}_{IJ}  \right) = 0, \label{so8deriv1}
\end{equation}
where the SO(8) gauge covariant derivative is defined as
\begin{equation}
 D^{\textup{SO}(8)}_{\mu} {w^{\pm}}_{i j}{}^{IJ} = \partial_{\mu} {w^{\pm}}_{i j}{}^{IJ} - 2 \sqrt{2} m_7 A_{\mu}{}^{K[I} {w^{\pm}}_{i j}{}^{J]K}.
 \end{equation}
Thence, we identify $A_{\mu}{}^{IJ}$ as the SO(8) gauge fields and $\sqrt{2} m_7$ 
as the SO(8) gauge coupling. Thus, the generalised vielbein postulate reduces to a 
particular component of the E$_{7(7)}$ Cartan equation with SO$(8)$ 
covariant derivatives, as claimed above.

While this part of the argument was already given in \cite{dWNconsis},
the SO(8) covariantisation on the other components of the generalised vielbein 
cannot be traced back to geometrical arguments of this type.
In Ref. \cite{dWNconsis} it is argued that equation \eqref{so8deriv1} in fact implies the E$_{7(7)}$ Cartan equation with SO$(8)$ covariant derivatives. However, here, by considering 
{\em all} of the generalised vielbein postulates, we can see that this equation 
follows directly upon compactification on $S^7.$ In other words, the rest of the generalised vielbein postulates give rise to the `missing' components of the Cartan 
equation in \eqref{CM}. We will show this in turn for each of the 
generalised vielbein postulates \eqref{cDB2}--\eqref{cDB4}.

The generalised vielbein postulate for $e_{mnAB},$ \eqref{cDB2}, becomes, 
after conversion to `flat' SU(8) indices using equation \eqref{Desu8trans2} and again using $\Do_m B_\mu{}^m=0$,
\begin{equation}
\cD_\mu e_{mn\,ij}  \,+\, 2 \Do_{[m} B_{|\mu|}{}^{p} e_{n]p\,ij} 
\,+\, 3 \Do_{[m} B_{|\mu| \, np]} e^{p}_{ij} 
\, +\,  \cQ_{\mu}^{k}{}_{[i} e_{mn\, j]k} \, + \, \cP_{\mu ijkl} e_{mn}{}^{kl} = 0.
\end{equation}
The new feature here is the presence of
the `magnetic'  vectors $B_{\mu mn},$ which according to equation \eqref{vec24d} 
could in principle lead to gauging of magnetic vector fields in the four-dimensional 
theory. However, note that the relations between $e_{mnAB},$ \eqref{gv24d}, and $B_{\mu mn},$ \eqref{vec24d}, and the four-dimensional fields are not made with respect to Killing vectors but via the tensor $K_{mn}^{IJ}$, which from the Killing spinor equation satisfies 
\begin{equation}
\Do_{p} K_{m n}{}^{IJ} = 2 m_7 \go_{p[m} K_{n]}{}^{ IJ}. \label{K2deriv}
\end{equation}
This immediately implies that 
\begin{equation}
\Do_{[m} B_{|\mu| np]} = 0. \label{dB2}
\end{equation}
Hence, {\em  the magnetic vector fields drop out of relation} \eqref{cDB2} in 
the $S^7$ reduction, thus ensuring that effectively only the 28 electric vectors
appear in the four-dimensional theory with their non-abelian interactions, while the
magnetic vectors all decouple!
Using equations \eqref{K1deriv} and \eqref{K2deriv}, 
the generalised vielbein postulate then further simplifies to 
\begin{equation}
 \partial_{\mu} e_{mn ij} + \frac{1}{12} m_7 A_{\mu}{}^{IJ} w^{-}{}_{ij}{}^{KL} \left(K^{p}{}_{[m}{}^{IJ} K_{n]p}{}^{KL} - K_{[m}{}^{IJ} K_{n]}{}^{KL} \right) + \cQ_{\mu}^{k}{}_{[i} e_{mnj]k} + \cP_{\mu ijkl} e_{mn}{}^{kl} = 0,
\end{equation}
which, using equation \eqref{gammareln2}, gives another component of the E$_{7(7)}$ Cartan equation:
\begin{equation}
 \label{so8deriv2}
 K_{m n}{}^{IJ} \left( D^{\textup{SO}(8)}_{\mu} {w^{-}}_{i j}{}^{IJ} + \cQ_{\mu}^{k}{}_{[i} {w^{-}}_{j]k}{}^{IJ} + \cP_{\mu ijkl} {w^{-}}^{kl}{}_{IJ}  \right) = 0.
\end{equation}

Next we consider the third equation, \eqref{cDB3}, which becomes, 
using $\Do_m B_\mu{}^m = 0$ and  equations \eqref{Desu8trans3} and \eqref{dB2},
\begin{align}
 &\partial_\mu e_{m_1 \dots m_5 ij} - B_{\mu}{}^{p} \Do_{p} e_{m_1 \dots m_5 ij} - 5 \Do_{[m_1} B_{|\mu|}{}^{p} e_{m_2 \dots m_5]p ij} - 6 \Do_{[m_1} B_{|\mu| m_2 \dots m_5 p]} e^{p}_{ij} 
 \notag \\[2mm]
 &\hspace{70mm} + \,\cQ_{\mu}^{k}{}_{[i} e_{m_1 \dots m_5 j]k}
 \, + \, \cP_{\mu ijkl} e_{m_1 \dots m_5}{}^{kl}=0. \label{gvp3}
\end{align}
In this case the reduction ans\"atze \eqref{vec34d} and \eqref{gv34d} not only contain tensors, rather than Killing vectors, but they also contain the potential for the volume-form on the round 7-sphere $\zetao,$ which is not globally defined. As we shall see below, these terms are not only crucial for obtaining the correct non-linear flux ans\"atze, 
but equally crucial in the reduction of the generalised vielbein postulate. 

Inserting the reduction ans\"atze for the generalised vielbeine and the vector fields and using equations \eqref{gamma52}, \eqref{K1deriv} and \eqref{K2deriv}, we obtain
\begin{align}
 B_{\mu}{}^{p} \Do_{p} e_{m_1 \dots m_5 ij} &=  \frac{1}{4\cdot 6!} A_{\mu}{}^{IJ} w^{+}{}_{ij}{}^{KL} K^{p IJ} \Big( m_7 \etao_{m_1 \dots m_5 p q} K^{q KL} + 6 \cdot 6! \Do_{p} \zetao_{m_1 \dots m_5 q} K^{q KL} \notag \\[2mm] 
 & \hspace{74mm} + 6 \cdot 6! m_7 \zetao_{m_1 \dots m_5 q} K^{q}{}_{p}{}^{KL} \Big), \\[3mm]
\Do_{[m_1} B_{|\mu| m_2 \dots m_5 p]} e^{p}_{ij} &= - \frac{1}{4\cdot 6!} A_{\mu}{}^{IJ} w^{+}{}_{ij}{}^{KL} K^{p KL} \Big( m_7 \etao_{m_1 \dots m_5 p q} K^{q IJ} - 6 \cdot 6! \Do_{[m_1} \zetao_{m_2 \dots m_5p]q} K^{q IJ} \notag \\
& \hspace{74mm} - 6 \cdot 6! m_7 \zetao_{q [m_1 \dots m_5} K^{q}{}_{p]}{}^{IJ} \Big).
\end{align}
Hence,
\begin{align}
 & B_{\mu}{}^{p} \Do_{p} e_{m_1 \dots m_5 ij} + 6 \Do_{[m_1} B_{|\mu| m_2 \dots m_5 p]} e^{p}_{ij} \notag \\[3mm]
 =& \frac{1}{4\cdot 6!} A_{\mu}{}^{IJ} w^{+}{}_{ij}{}^{KL} \Big(7 m_7 \etao_{m_1 \dots m_5 p q}  K^{p IJ} K^{q KL} + 6 \cdot 7!  \Do_{[p} \zetao_{m_1 \dots m_5 q]} K^{p IJ} K^{q KL} \notag \\[1mm]
 & \hspace{35mm} + 6 \cdot 6! m_7 \zetao_{m_1 \dots m_5 q} K^{p IJ} K^{q}{}_{p}{}^{KL} + 36 \cdot 6! m_7 \zetao_{q [m_1 \dots m_5} K^{q}{}_{p]}{}^{IJ} K^{p KL} \Big), \notag \\[2mm] 
 =& \frac{1}{4\cdot 6!} m_7 A_{\mu}{}^{IJ} w^{+}{}_{ij}{}^{KL}  \Big(\etao_{m_1 \dots m_5 p q}  K^{p IJ} K^{q KL} + 48 \cdot 6! \zetao_{m_1 \dots m_5 q} \delta^{IK} K^{qJL} \notag \\
 & \hspace{85mm} + 30 \cdot 6! \zetao_{p q [m_1 \dots m_4} K^{q}{}_{m_5]}{}^{IJ} K^{p KL} \Big), \label{be5b5e}
\end{align}
where in the second equality above we have used equations \eqref{7volpot} and \eqref{gammareln1}.
A straightforward substitution of the ans\"atze \eqref{vec14d} and \eqref{gv34d} also gives
\begin{align}
5 \Do_{[m_1} B_{|\mu|}{}^{p} e_{m_2 \dots m_5]p ij} & = \frac{5 m_7}{8 \cdot 6!} A_{\mu}{}^{IJ} w^{+}{}_{ij}{}^{KL} K^{p}{}_{[m_1}{}^{IJ} \Big(\etao_{m_2 \dots m_5] pqr} K^{qrKL} + 12 \cdot 6! \zetao_{m_2 \dots m_5] pq} K^{q KL} \Big). \label{be5}
\end{align}
Now, using the identity
\begin{equation}
 5  K^{p}{}_{[m_1}{}^{IJ} \etao_{m_2 \dots m_5] pqr} K^{qrKL} = 2  K^{p}{}_{r}{}^{IJ} \etao_{m_1 \dots m_5 pq}K^{qrKL}
\end{equation}
and equation \eqref{gammareln2}, the first terms on the right hand side of equations \eqref{be5b5e} and \eqref{be5} precisely combine to give
\begin{equation}
 \frac{1}{6!} m_7 A_{\mu}{}^{IJ} w^{+}{}_{ij}{}^{KL} \etao_{m_1 \dots m_5 p q} \delta^{IK} K^{pqJL} = - \frac{2}{6!} m_7 A_{\mu}{}^{IJ} w^{+}{}_{ij}{}^{KL} \delta^{IK} K_{m_1 \dots m_5}{}^{JL}.
\end{equation}
Moreover, the third term on the right hand side of equation \eqref{be5b5e} cancels the second term on the right hand side of equation \eqref{be5}. Therefore, in all, equation \eqref{gvp3} simplifies to 
\begin{equation}
 \left(K_{m_1 \dots m_5}{}^{IJ} - 6 \cdot 6! \zetao_{m_1 \dots m_5 p} K^{pIJ} \right) \left[ D^{\textup{SO}(8)}_{\mu} {w^{+}}_{i j}{}^{IJ} + \cQ_{\mu}^{k}{}_{[i} {w^{+}}_{j]k}{}^{IJ} + \cP_{\mu ijkl} {w^{+}}^{kl}{}_{IJ} \right]= 0. 
\end{equation}
Using equations \eqref{so8deriv1} and \eqref{gamma52}, the above equation implies 
\begin{equation}
\label{so8deriv3}
 K^{m n IJ} \left( D^{\textup{SO}(8)}_{\mu} {w^{+}}_{i j}{}^{IJ} + \cQ_{\mu}^{k}{}_{[i} {w^{+}}_{j]k}{}^{IJ} + \cP_{\mu ijkl} {w^{+}}^{kl}{}_{IJ}  \right) = 0.
\end{equation}

Finally, we consider the generalised vielbein postulate for $e_{m_1 \dots m_7,n},$ which using the same equations as before, simplifies to
\begin{align}
 &\partial_\mu e_{m_1 \dots m_7, n ij} -  B_{\mu}{}^{p} \Do_{p} e_{m_1 \dots m_7, n ij} - \partial_{n} B_{\mu}{}^{p} e_{m_1 \dots m_7, p ij} - 2 \partial_{[m_1} B_{|\mu| m_2 \dots m_6} e_{m_7] n ij} \notag \\[2mm] 
& \hspace{80mm} + \cQ_{\mu}^{k}{}_{[i} e_{m_1 \dots m_7, n j]k} + \cP_{\mu ijkl} e_{m_1 \dots m_7, n}{}^{kl} =0.
\label{gvp4}
\end{align}
A similar calculation to the one outlined above for the $e_{m_1 \dots m_5 AB}$ gives 
\begin{align}
 B_{\mu}{}^{p} \Do_{p} e_{m_1 \dots m_7, n ij} &= - \frac{\sqrt{2}}{2 \cdot 9!} A_{\mu}{}^{IJ} w^{-}{}_{ij}{}^{KL} \Big( 5 m_7 \etao_{m_1 \dots m_7} K^{p IJ} K_{pn}^{KL} + 36 \cdot 7! \Do_{[m_1|} \zetao_{p|m_2 \dots m_6} K^{p IJ} K_{m_7]n}^{KL} \notag \\[2mm] 
 & \hspace{50mm} + 6 \cdot 7! m_7 \zetao_{[m_1 \dots m_6} \left( K_{m_7]}{}^{IJ} K_{n}^{KL} - K_{m_7]}{}^{KL} K_{n}^{IJ} \right) \Big),
 \end{align}
\begin{align}
\partial_{n} B_{\mu}{}^{p} e_{m_1 \dots m_7, p ij} & = - \frac{\sqrt{2}}{2 \cdot 9!} m_7 A_{\mu}{}^{IJ} w^{-}{}_{ij}{}^{KL} K^{p}{}_{n}{}^{IJ} \Big(\etao_{m_1 \dots m_7} K_{p}{}^{KL} + 6 \cdot 7! \zetao_{[m_1 \dots m_6} K_{m_7]p}{}^{KL} \Big),
\end{align}
 \begin{align}
2 \partial_{[m_1} B_{|\mu| m_2 \dots m_6} e_{m_7] n ij} &= \frac{\sqrt{2}}{2 \cdot 9!} A_{\mu}{}^{IJ} w^{-}{}_{ij}{}^{KL} \Big( 6 m_7 \etao_{m_1 \dots m_7} K^{p IJ} K_{pn}^{KL} \notag \\
&\hspace{43mm} + 36 \cdot 7! \Do_{[m_1|} \zetao_{p|m_2 \dots m_6} K^{p IJ} K_{m_7]n}^{KL} \notag \\[2mm]
& \hspace{54mm} + 6 \cdot 7! m_7 \zetao_{[m_1 \dots m_6} K_{m_7]p}{}^{IJ} K^{p}{}_{n}{}^{KL} \Big),
\end{align}
where we have used equations \eqref{K1deriv}, \eqref{7volpot}, \eqref{K2deriv}, \eqref{gamma52} and antisymmetrisations over eight indices, which vanish, to simplify the expressions above. It is now simple to verify, using equations \eqref{gammareln1} and \eqref{gammareln2}, that
\begin{align}
 & B_{\mu}{}^{p} \Do_{p} e_{m_1 \dots m_7, n ij} + \partial_{n} B_{\mu}{}^{p} e_{m_1 \dots m_7, p ij} + 2 \partial_{[m_1} B_{|\mu| m_2 \dots m_6} e_{m_7] n ij} \notag \\[3mm]
& \hspace{45mm} =  \frac{4 \sqrt{2}}{9!} m_7 A_{\mu}{}^{KI} w^{-}{}_{ij}{}^{JK} \Big( \etao_{m_1 \dots m_7} K_{n}^{IJ}  + 6 \cdot 7! \zetao_{[m_1 \dots m_6} K_{m_7]n}{}^{IJ} \Big).
\end{align}
Hence equation \eqref{gvp4} reduces to 
\begin{equation}
 \label{so8deriv4}
K_{m}{}^{IJ} \left( D^{\textup{SO}(8)}_{\mu} {w^{-}}_{i j}{}^{IJ} + \cQ_{\mu}^{k}{}_{[i} {w^{-}}_{j]k}{}^{IJ} + \cP_{\mu ijkl} {w^{-}}^{kl}{}_{IJ}  \right) = 0,
\end{equation}
where we have used equation \eqref{so8deriv2} to eliminate the expression proportional to $\zetao$.

To sum up: in the reduction of \eqref{cDB1}, the SO(8) gauge covariant derivative 
arose from the geometrical properties of Killing vectors on $S^7.$ However, for the other generalised vielbeine the emergence of the  SO(8) gauge covariant derivative is not so direct. Indeed, the reduction of the other generalised vielbein postulates is particularly novel given that the postulates \eqref{cDB2}--\eqref{cDB4} contain fields ($B_{\mu mn}$ and $B_{\mu m_1 \dots m_5}$) for which the identification with the four-dimensional vector fields is not made with the $S^{7}$ Killing vectors, but with more general structures on the 7-sphere. We stress again that in the derivation of the last two equations, \eqref{so8deriv3} and \eqref{so8deriv4}, the $\zetao$ terms in the ans\"atze are crucial for obtaining the SO(8) gauge covariant terms. Therefore, the SO(8) gauge covariant derivatives 
emerge, not in spite of but {\em because of} these more general structures. 

The results that we have obtained from the reduction of the generalised vielbein postulates, equations \eqref{so8deriv1}, \eqref{so8deriv2}, \eqref{so8deriv3} and \eqref{so8deriv4}, can be summarised as 
\begin{align} \label{so8derivK1}
 & K^{a IJ} \left( D^{\textup{SO}(8)}_{\mu} {w^{\pm}}_{i j}{}^{IJ} + \cQ_{\mu}^{k}{}_{[i} {w^{\pm}}_{j]k}{}^{IJ} + \cP_{\mu ijkl} {w^{\pm}}^{kl}{}_{IJ}  \right) = 0, \\[8pt]\label{so8derivK2}
& K^{ab IJ} \left( D^{\textup{SO}(8)}_{\mu} {w^{\pm}}_{i j}{}^{IJ} + \cQ_{\mu}^{k}{}_{[i} {w^{\pm}}_{j]k}{}^{IJ} + \cP_{\mu ijkl} {w^{\pm}}^{kl}{}_{IJ}  \right) = 0.
\end{align}
Since $K^{a IJ}$ and $K^{ab IJ}$ form a basis of antisymmetric $28 \times 28$ matrices, these equations are equivalent to 
\begin{equation}
 D^{\textup{SO}(8)}_{\mu} {\mathcal{V}}_{i j}{}^{IJ} + \cQ_{\mu}^{k}{}_{[i} {\mathcal{V}}_{j]k}{}^{IJ} + \cP_{\mu ijkl} {\mathcal{V}}^{kl}{}_{IJ} =0,
\end{equation}
where $\mathcal{V}(x)$ is the E$_{7(7)}/$SU(8) coset element parametrised by the scalar 
fields. In Ref. \cite{dWNconsis} this equation was argued, somewhat indirectly,
to hold solely on the basis
of the generalised vielbein postulate for $e^{m}_{AB}$, equation \eqref{cDB1}. Here 
we see that it naturally follows from the full set of generalised vielbein postulates. 

In summary, we find that, in the case of the $S^7$ compactification both $B_{\mu}{}^{m}$ 
and $B_{\mu m_1 \dots m_5}$ contribute to the electric vector fields, while the magnetic 
vector fields drop out of the expressions.
Indeed, from equations \eqref{vec14d} and \eqref{vec34d} 
we see that this is natural because $B_{\mu}{}^{m}$ and $B_{\mu m_1 \dots m_5}$ 
project onto different SO(8) components of the electric vector field $A_{\mu}{}^{IJ}.$

\section{A first test of the non-linear six-form ansatz} \label{sec:test}

In this section we check the consistency of the relations derived in section \ref{sec:NLA}, in particular the non-linear ansatz for the dual six-form using the relatively simple, yet non-trivial SO$(7)^\pm$ invariant solutions of gauged supergravity \cite{Warner83} for which the higher-dimensional solutions are known \cite{dWNso7soln, englert}.  For the convenience of the reader, 
we give a brief description of these solutions from both a four and higher-dimensional 
perspective in appendix \ref{app:so7}. The non-linear metric and flux ans\"atze, 
equations \eqref{metansatz} and \eqref{potans}, respectively, have been subjected 
to some very non-trivial tests, which they have passed with remarkable success,
most recently in \cite{GGN} (where references to earlier work can also be found).
In particular, these ans\"atze correctly reproduce the SO$(7)^\pm$ invariant solutions \cite{dWNW, NP, GGN}.  Therefore, let us consider the non-linear ansatz, equation \eqref{6formans}, for the dual six-form potential of the SO$(7)^\pm$ invariant solutions.

Using the following E$_{7(7)}$ properties satisfied by $u^{ij}{}_{KL}$ and $v^{ij KL}$ \cite{dWNn8}
\begin{align}
 u^{ij}{}_{IJ} u_{ij}{}^{KL} - v_{ijIJ} v^{ijKL} &= \delta^{KL}_{IJ},\label{e7reln1} \\
 u^{ij}{}_{IJ} v_{ijKL} - v_{ijIJ} u^{ij}{}_{KL} &= 0, \label{e7reln2}
\end{align}
one can show that
\begin{equation}
 ({{u_{ij}}^{IJ}}+v_{ijIJ}) ( u^{ij}{}_{KL} + v^{ij KL}) = \delta^{IJ}_{KL} + 2\, v_{IJMN} v^{KLMN} + 2\, \textup{Re}({u_{MN}}^{IJ} v^{MNKL}).
\end{equation}
Now, using the form of $u^{ij}{}_{KL}$ and $v^{ij KL}$ for the SO$(7)^\pm$ invariant solutions, given in appendix \ref{app:so7}, and the following identifies satisfied by $C_{\pm}^{IJKL}$ \cite{dWNso7soln, parallel}
\begin{align}
 C_{+}^{IJMN} C_{+}^{MNKL} &= 12 \delta^{IJ}_{KL} + 4 C_{+}^{IJKL}, \label{C+C+} \\
 C_{-}^{IJMN} C_{-}^{MNKL} &= 12 \delta^{IJ}_{KL} - 4 C_{-}^{IJKL} \label{C-C-},
\end{align}
the above equation reduces to
\begin{equation}
 ({{u_{ij}}^{IJ}}+v_{ijIJ}) ( u^{ij}{}_{KL} + v^{ij KL}) = (c^3 + \epsilon\, s^3) \delta^{IJ}_{KL} +\frac{1}{2} \epsilon\, cs(c+s) C_+^{IJKL} -\frac{1}{2}(1-\epsilon)\, cs^2 C_{-}^{IJKL},
\end{equation}
where
\begin{equation}
  \epsilon = \begin{cases}
                     1 & \quad \textup{SO}(7)^+ \\[1mm]
		     0 & \quad \textup{SO}(7)^-
                      \end{cases}.
\end{equation}
Dualising $\Gamma_{m_1 \ldots m_5}$ and using the identities satisfied by the contraction of $C_{\pm}^{IJKL}$ with $K^{m IJ}$ and $K^{mn IJ}$ (see Ref.~\cite{GGN}) gives
\begin{align}
&A_{n m_1 \dots m_5} -  \frac{\sqrt{2}}{4} A_{n[m_1 m_2} A_{m_3 m_4 m_5]} \notag \\[4mm]
&\hspace{8mm}=\frac{\sqrt{2}}{4\cdot 6!}\, \Delta\, g_{np}\, \Bigg\lbrace 12\cdot6! (c^3 + \epsilon s^3) {\zetao_{m_1 \ldots m_5}}^{p} \notag \\[2mm]
&\hspace{35mm}- \frac{cs(c+s)}{3} \epsilon \left[{{\etao_{m_1\ldots m_5}}^{p}}_{q} \xi^q + 6\cdot6! \zetao_{m_1 \ldots m_5 q}\left((3+\xi) \go^{pq} - (21+\xi)\hat{\xi}^m \hat{\xi}^n\right) \right] \notag \\[2mm]
&\hspace{35mm}-(1-\epsilon) cs^2 \etao_{m_1 \ldots m_5 rs} \So^{prs} \Bigg\rbrace.
\label{A6so7pm}
\end{align}
To compare this result, obtained from the uplift formula, with the results directly
obtained by solving the $D=11$ field equations, let us 
first consider the SO$(7)^+$ invariant solution.  Using 
\begin{equation}
 cs(c+s)=\gamma^{1/2}/5, \qquad c^3+s^3=2\gamma^{1/2}/5
\end{equation}
and noting that $A_{mnp}=0$ for this solution, equation \eqref{A6so7pm} reduces to
\begin{equation}
 A_{m_1 \dots m_6} = \frac{\sqrt{2}}{2\cdot6!}\, \frac{1}{9-\xi}\, \etao_{m_1 \dots m_6 p} \xi^p -3\sqrt{2} \zetao_{m_1 \dots m_6},
\end{equation}
where we have used the form of the metric $g_{mn}$ given in equation \eqref{gso7+}.  
Now, taking the exterior derivative of this equation gives
\begin{equation}
 7! \Do_{[m_1}A_{m_2 \ldots m_7]} = -\frac{180\sqrt{2}}{(9-\xi)^2} m_7\; \etao_{m_1 \ldots m_7},
\end{equation}
which agrees precisely with the expression found by dualising the Freund-Rubin field strength for the value of $\gamma$ set by the non-linear ans\"atze, see equation \eqref{A611dso7+}.

Next, consider the SO$(7)^-$ invariant solution.  Using the expressions for $A_{mnp}$ and $g_{mn}$ in appendix \ref{app:so7}, it is simple to show that equation \eqref{A6so7pm} reduces to
\begin{equation}
 A_{m_1 \ldots m_6} = -3\sqrt{2} \zetao_{m_1 \ldots m_6} + \frac{\sqrt{2}}{16\cdot6!}\gamma^{-1/3} {\So^{pq}}_{[m_1} \etao_{m_2 \ldots m_6] pq}.
\end{equation}
However, the second term on the right hand side vanishes
by the Schouten identity and the tracelessness of the torsion. Hence,
\begin{equation}
 A_{m_1 \ldots m_6} = -3\sqrt{2} \zetao_{m_1 \ldots m_6},
\end{equation}
which agrees with the expression for $A_{m_1 \ldots m_6}$ given in equation 
\eqref{A611dso7-} for the particular value of $\gamma$ set by the ans\"atze.

\section{Outlook: magnetic vectors and the embedding tensor} \label{sec:outlook}

Having established the full consistency of all equations for the $S^7$ compactification
we now return to the most remarkable feature of the vielbein postulate equations 
\eqref{cDB1}--\eqref{cDB4}, namely the fact that they simultaneously involve 
the Kaluza-Klein vectors, the `non-geometric vectors' coming from the three-form
field, and the $D=11$ dual vector fields. In principle, it is therefore clear that both 
`electric' vector fields coming from the reduction of $B_{\mu}{}^{m}$ and 
$B_{\mu m_1 \dots m_5},$ and `magnetic' vector fields, coming from the 
reduction of $B_{\mu mn}$,  can be gauged. This is a feature that our construction 
shares with the embedding tensor formalism as applied to gaugings of maximal
supergravity in four dimensions \cite{NScomgauge3,dWSTlag, dWSTmax4}. 
There as well, one initially works with the full set of 56 electric and magnetic vector fields, replacing the Cartan equations \eqref{CM} with the more general ansatz
\begin{equation}\label{CM1}
\cV^{-1}(x) \Big[ \partial_\mu \, -\, g A_\mu^{IJ}(x) \Theta_{IJ\,\cA} \cY^\cA
 -\, g A_{\mu\,IJ}(x) \Theta^{IJ}{}_{\cA} \cY^\cA\Big] \cV(x)
= Q_\mu(x) + P_\mu (x),
\end{equation}
where $\cY^\cA$ ($\cA= 1, \dots, 133$) are the generators of E$_{7(7)}$, 
$\big( \Theta_{IJ\,\cA} , \Theta^{IJ}{}_\cA\big)$ is the embedding tensor, and
$A_\mu^{IJ}$ and $A_{\mu IJ}$ are the electric and magnetic vectors introduced
in \eqref{susyev} and \eqref{susymv}, respectively.
The embedding tensor thus transforms in the product ${\bf 56} \otimes {\bf 133}$,
but a  consistent gauging with 28 propagating gauge fields exists only when $\Theta$ 
restricts to the $\bf{912}$ representation of E$_{7(7)}$  in this product 
\cite{NScomgauge3,dWSTlag, dWSTmax4} 
(and in addition satisfies a quadratic identity). The choice of embedding tensor
not only determines the gauge group, but also decides which 28 vector
fields out of the initial 56 vectors become propagating non-abelian vectors. 
Consequently, studying the vielbein 
equations \eqref{cDB1}---\eqref{cDB4} in parallel with \eqref{CM1} should thus 
enable one to understand the embedding tensor and its relation to any particular
compactification directly from the eleven-dimensional perspective. Although we will leave the full exploration of these possibilities to future work, we conclude with some comments.

In the $S^7$ compactification considered in section \ref{sec:gvp}, the SO$(8)$ gauge covariant derivative term comes from the following terms in the generalised vielbein postulates:
\begin{equation}
 \partial_{m} B_{\mu}{}^{n} \qquad \textup{and} \qquad \partial_{[m_1} B_{|\mu| m_2 \dots m_6]}.
\end{equation}
In particular, terms of the form 
\begin{equation} \label{term0}
 \partial_{m} B_{\mu}{}^{m}  \qquad \textup{and} \qquad \partial_{[m} B_{|\mu|np]}
\end{equation}
do not contribute. The first expression above vanishes because the Kaluza-Klein ansatz for the graviphoton is given by $S^7$ Killing vectors, which are divergence-free; while the second expression vanishes because of the form of the Kaluza-Klein ansatz for $B_{\mu mn}$, equation \eqref{vec24d}, and properties of $S^7$ Killing spinors. 

A natural question to ask is whether one can find examples of compactifications where the expressions \eqref{term0}, that vanish for the $S^7$ reduction, contribute to the four-dimensional Cartan equation \eqref{CM1}. For example, while the first expression vanishes for Killing vectors, it is non-zero for conformal Killing vectors, which also form a simple Lie algebra. An interesting question is whether one can carry out more general reductions of this type.      

Furthermore, a particularly interesting class of gaugings
to investigate in this context are the Scherk-Schwarz compactifications \cite{ScherkSchwarz} and twisted 7-torus flux compactifications \cite{ttf1, ttf2, Mtwistedtorus, ttf3, ttf4}, which lead to various gaugings in four dimensions (see \cite{dWSTmax4} for a review of known gaugings in four dimensions). While the original Scherk-Schwarz reductions on flat groups are known to lead to electric gaugings \cite{ADFL}, flux compactifications provide examples where both electric and magnetic vector fields contribute to the gauging \cite{Mtwistedtorus}.

A study of the generalised vielbein postulates may also shed light on the higher-dimensional origins of the recently discovered continuous family of inequivalent maximal SO$(8)$ gauged supergravities \cite{DIT}. While the original SO$(8)$ gauged supergravity \cite{dWNn8} in the SL$(8,\mathbb{R})$ symplectic frame only contains electric gaugings, there is a deformation that allows both electric and magnetic gaugings in the aforementioned symplectic frame. For a given range of the angle of rotation between gaugings of electric and magnetic vector fields the theory is inequivalent to the original theory. While $D=11$ supergravity apparently cannot explain 
the existence of these new supergravities, with the framework presented here 
it is possible to investigate whether $D=11$ supergravity admits an analogous deformation 
that rotates $B_{\mu}{}^{\tM \tN}$ and $B_{\mu \tM \tN}$, defined in \eqref{Bdef}, into each other
and that would be implemented by a rotation on the 56-bein \eqref{56bein} in complete  
analogy with the $D=4$ theory \cite{dWN13}. The $S^7$ compactification of these putative theories would then give rise to the  magnetic gaugings in the deformed 
theories found in \cite{DIT}.

\newpage
\appendix

\section{SO$(7)^{\pm}$ invariant solutions} \label{app:so7}

In this appendix, we summarise the SO$(7)^{\pm}$ invariant stationary points of maximal gauged supergravity \cite{Warner83} and their respective eleven-dimensional counterpart solutions \cite{dWNso7soln, englert}.  The non-linear metric and flux ans\"atze have been confirmed for these solutions in Ref.~\cite{dWNW, GGN}.  Much of the necessary information regarding these solutions is explained in Ref.~\cite{GGN} and in particular its appendix A.  Therefore, for brevity, we refer the reader there for the definitions of the relevant structures and content ourselves here with a list of the most important properties of these solutions that will be relevant for the calculations in section \ref{sec:test}.

The scalar profile for the SO$(7)^+$ invariant stationary point is given by \cite{parallel,dWNW}
\begin{align}
 {u^{IJ}}_{KL} = p^3 \delta^{IJ}_{KL} + \textstyle{\frac{1}{2}} p q^2 C_+^{IJKL}, \label{so7+u} \\[2mm]
 v^{IJKL} = q^3 \delta^{IJ}_{KL} + \textstyle{\frac{1}{2}} p^2 q C_+^{IJKL}, \label{so7+v}
\end{align}
where constants $p$ and $q$ are such that \cite{Warner83}
\begin{align}
 c^2&=(p^2+q^2)^2= \frac{1}{2} (3/\sqrt{5} +1), \\
 s^2&=(2pq)^2 = \frac{1}{2} (3/\sqrt{5} -1).
\end{align}

The eleven-dimensional solution is of the form \cite{dWNso7soln}
\begin{eqnarray}
 g_{MN} &=&\gamma^{7/18}\, 30^{-2/3} (9-\xi)^{2/3}  \left(\etao_{\mu \nu},\ \frac{\gamma^{-1/2}}{9-\xi} \left[30\, \go_{mn} - (21+\xi)\, \hat{\xi}_m \hat{\xi}_n \right]\right), \quad
 \label{gso7+} \\[2mm] 
F_{MNPQ}&=&\left(\frac{\sqrt{6}}{3} i \, m_7 \, \gamma^{5/6} \etao_{\mu \nu \rho \sigma},\ 0 \right),
\end{eqnarray}
where $\gamma$ is an arbitrary constant, which takes the value
\begin{equation}
 \gamma=5^{3/2}
\end{equation}
when the solution is constructed via the non-linear ans\"atze \cite{GGN}.  Note that the determinant of the siebenbein
\begin{equation}
 \Delta=\textup{det} \, ({e_m}^a) = \sqrt{\textup{det}(g_{mn})}=\gamma^{-7/18}\, 30^{2/3} (9-\xi)^{-2/3}.
\end{equation}

In addition, due to the existence of the Freund-Rubin term, the dual potential $A_{(6)}$ is non-zero and of the form
\begin{equation} \label{A611dso7+}
 7! D_{[M_1} A_{M_2 \ldots M_7]} = \begin{cases}
                     - 180\sqrt{10} \, m_7\, \gamma^{-1/3} (9-\xi)^{-2} \etao_{m_1 \ldots m_7} &
                     \quad  [m_1 \ldots m_7] \\[1mm]
			0 & \quad  \text{otherwise}
                      \end{cases}.
\end{equation}

The scalar profile of the SO$(7)^-$ invariant stationary point of maximal supergravity is of the form \cite{parallel,dWNW}
\begin{align}
 {u^{IJ}}_{KL} = p^3 \delta^{IJ}_{KL} - \textstyle{\frac{1}{2}} p q^2 C_-^{IJKL}, \\[2mm]
 v^{IJKL} = i q^3 \delta^{IJ}_{KL} - \textstyle{\frac{1}{2}} i p^2 q C_-^{IJKL},
\end{align}
where constants $c$ and $s$, related to $p$ and $q$, as above take the values \cite{Warner83}
\begin{equation}
  c^2=\frac{5}{4}, \qquad s^2=\frac{1}{4}.
\end{equation}

The eleven-dimensional solution is of the form \cite{englert}
\begin{eqnarray}
 g_{MN} &=&\gamma^{7/18}\left(\etao_{\mu \nu},\ \gamma^{-1/2} \go_{mn}\right), \quad
 \\[1mm] 
F_{MNPQ}&=&\left(2 \sqrt{2} i \, m_7 \, \gamma^{5/6} \etao_{\mu \nu \rho \sigma},\ \frac{\sqrt{2}}{6} m_7 \, \gamma^{-1/6} \, \etao_{mnpqrst} \So^{rst}\right),
\end{eqnarray}
and in particular,
\begin{equation}
 A_{M N P} = \begin{cases}
                     2\sqrt{2}\, i \gamma^{5/6} \zetao_{\mu \nu \rho} & \quad  [\mu \nu \rho] \\[1mm]
                     \frac{\sqrt{2}}{4!} \gamma^{-1/6} \So_{mnp} & \quad  [mnp] \\[1mm]
			0 & \quad  \text{otherwise}
                      \end{cases},
\end{equation}
where $\zetao_{\mu \nu \rho}$ is the potential for the Freund-Rubin field strength
\begin{equation}
 4! \, \partial_{[\mu} \zetao_{\nu \rho \sigma]} = m_7 \, \etao_{\mu \nu \rho \sigma}.
\end{equation}
As before, $\gamma$ is an arbitrary constant that is fixed by the non-linear ans\"atze to take the value \cite{GGN}
\begin{equation}
 \gamma^{1/3}=5/4.
\end{equation}
Furthermore,
\begin{equation}
  \Delta=\textup{det} \, ({e_m}^a) = \sqrt{\textup{det}(g_{mn})}=\gamma^{-7/18}.
\end{equation}
The six-form potential for this solution is of the form \cite{GGN13}
\begin{equation}   \label{A611dso7-}
 A_{M_1 \ldots M_6} = \begin{cases}
                       \frac{\sqrt{2}}{12}i \, \gamma^{2/3} \zetao_{\mu \nu \rho} \So_{mnp} & 
                      \quad  [\mu \nu \rho mnp] \\[1mm]
                     -\frac{15\sqrt{2}}{4} \gamma^{-1/3} \zetao_{m_1 \ldots m_6} &
                     \quad  [m_1 \ldots m_6] \\[1mm]
			0 & \quad  \text{otherwise}
                      \end{cases},
\end{equation}
where $\zetao_{m_1 \ldots m_6}$ is defined in equation \eqref{7volpot}.

\section{Useful identities} \label{app:iden}
 We list some useful identities satisfied by 7-dimensional $\Gamma$-matrices. These identities already appear in Refs. \cite{so(8)} and \cite{dWNsu8}.
 \begin{align}
  \Gamma^{a_1 \dots a_7} &= - i \epsilon^{a_1 \dots a_7}, \label{gamma70} \\
  \Gamma^{a_1 \dots a_6} &= - i \epsilon^{a_1 \dots a_6 b} \Gamma^{b}, \label{gamma61} \\
  \Gamma^{a_1 \dots a_5} &=  \frac{i}{2} \epsilon^{a_1 \dots a_5 b c} \Gamma^{bc}, \label{gamma52}\\
  \Gamma^{a_1 \dots a_4} &=  \frac{i}{3!} \epsilon^{a_1 \dots a_4 b c d} \Gamma^{bcd}, \label{gamma43}
 \end{align}
\begin{gather}
 \Gamma^{a b}_{AB} \Gamma^{b}_{CD} - \Gamma^{b}_{AB} \Gamma^{a b}_{CD}  =  8 \delta_{[C|[A} \Gamma^{a}_{B]|D]}, \label{gammareln1} \\
 \Gamma^{c[a}_{AB} \Gamma^{b]c}_{CD} + \Gamma^{[a}_{AB} \Gamma^{b]}_{CD} = 4 \delta_{[C|[A} \Gamma^{ab}_{B]|D]}.\label{gammareln2}
\end{gather}
These identities are exactly identities (A.1), (A.6) and (A.7) in the appendix A of \cite{dWNsu8}.

\newpage

\bibliography{gv1}
\bibliographystyle{utphys}
\end{document}